\documentclass{ieeeaccess}
\usepackage{amsthm}
\usepackage{amssymb}
\usepackage{mathptmx}
\usepackage{array}
\newcolumntype{P}[1]{>{\centering\arraybackslash}p{#1}}
\usepackage[cmex10]{amsmath}
\usepackage{amsfonts}
\usepackage{caption}
\usepackage{comment}
\usepackage{xcolor}
\usepackage{soul}
\usepackage[boxed,ruled,vlined,linesnumbered,commentsnumbered, noresetcount]{algorithm2e}
\usepackage{algorithmic}
\usepackage{setspace}
\usepackage{mathtools}
\usepackage{cite}
\usepackage{seqsplit}

\usepackage{flushend}

\usepackage[utf8x]{inputenc}



\usepackage{subcaption}

\makeatletter
\newcommand{\biggg}{\bBigg@{3}}
\newcommand{\vast}{\bBigg@{4}}
\newcommand{\Vast}{\bBigg@{5}}
\makeatother

\usepackage{color}
\usepackage{array}
\usepackage{threeparttable}

\usepackage{pifont}% http://ctan.org/pkg/pifont
\newcommand{\cmark}{\ding{51}}%
\newcommand{\xmark}{\ding{55}}%

\usepackage{textcomp}

\def\BibTeX{{\rm B\kern-.05em{\sc i\kern-.025em b}\kern-.08em
    T\kern-.1667em\lower.7ex\hbox{E}\kern-.125emX}}

\begin{document}
\history{Date of publication xxxx 00, 0000, date of current version xxxx 00, 0000.}
\doi{10.1109/ACCESS.2017.DOI}

\title{Machine Learning in Event-Triggered  Control: Recent Advances and Open Issues}
\author{\uppercase{Leila Sedghi}\authorrefmark{1},
%\IEEEmembership{Fellow, IEEE},
\uppercase{Zohaib Ijaz}\authorrefmark{1}, and \uppercase{Md. Noor-A-Rahim}\authorrefmark{1}, and \uppercase{Kritchai Witheephanich}\authorrefmark{2}, and \uppercase{Dirk Pesch}\authorrefmark{1}  }
\address[1]{ School of Computer Science \& IT, University College Cork,  Ireland 
(E-mail: {leila.sedghi@cs.ucc.ie, zohaib.ijaz@cs.ucc.ie, m.rahim@cs.ucc.ie \text{and} d.pesch@cs.ucc.ie})}
\address[2]{Department  of Electrical \& Electronic Engineering, Munster Technological University, Ireland  (E-mail:  Kritchai.Witheephanich@mtu.ie}

\tfootnote{This publication has emanated from research conducted with the financial support of Science Foundation Ireland under Grant number 16/RC/3918.}

\markboth
{Leila S. \headeretal: Machine Learning in Event-Triggered  Control}
{Leila S. \headeretal: Machine Learning in Event-Triggered  Control}

\corresp{Corresponding authors: Leila Sedghi (e-mail: leila.sedghi@cs.ucc.ie).}

\begin{abstract}
Networked control systems have gained considerable attention over the last decade as a result of the trend towards decentralised control applications and the emergence of cyber-physical system applications. However, real-world wireless networked control systems suffer from limited communication bandwidths, reliability issues, and a lack of awareness of network dynamics due to the complex nature of wireless networks. Combining machine learning and event-triggered control has the potential to alleviate some of these issues. For example, machine learning can be used to overcome the problem of a lack of network models by learning system behavior or adapting to dynamically changing models by continuously learning model dynamics. Event-triggered control can help to conserve communication bandwidth by transmitting control information only when necessary or when resources are available. The purpose of this article is to conduct a review of the literature on the use of machine learning in combination with event-triggered control. Machine learning techniques such as statistical learning, neural networks, and reinforcement learning-based approaches such as deep reinforcement learning are being investigated in combination with event-triggered control. We discuss how these learning algorithms can be used for different applications depending on the purpose of the machine learning use. Following the review and discussion of the literature, we highlight open research questions and challenges associated with machine learning-based event-triggered control and suggest potential solutions. 
\end{abstract}


\titlepgskip=-15pt

\maketitle

\section{Introduction}
\label{introduction}
Networked Control Systems (NCSs) have recently sparked a lot of interest as they provide solutions to a variety of technical problems in areas such as automotive systems, process control, smart manufacturing, smart grids, and autonomous driving. An NCS can be divided into two components: a cyber network and a physical component. This makes NCS a class of cyber-physical systems (CPSs) that are distributed across a network. %The cyber network is composed of intelligent network systems equipped with sensors, processors, and actuators that collect data from various agents (an agent can be any object connected to a system to form an NCS) and physical components (where the agent is physically connected to the system) in order to ensure real-time performance guarantees

\begin{figure} [t]
    \centering
   \includegraphics[width=\linewidth]{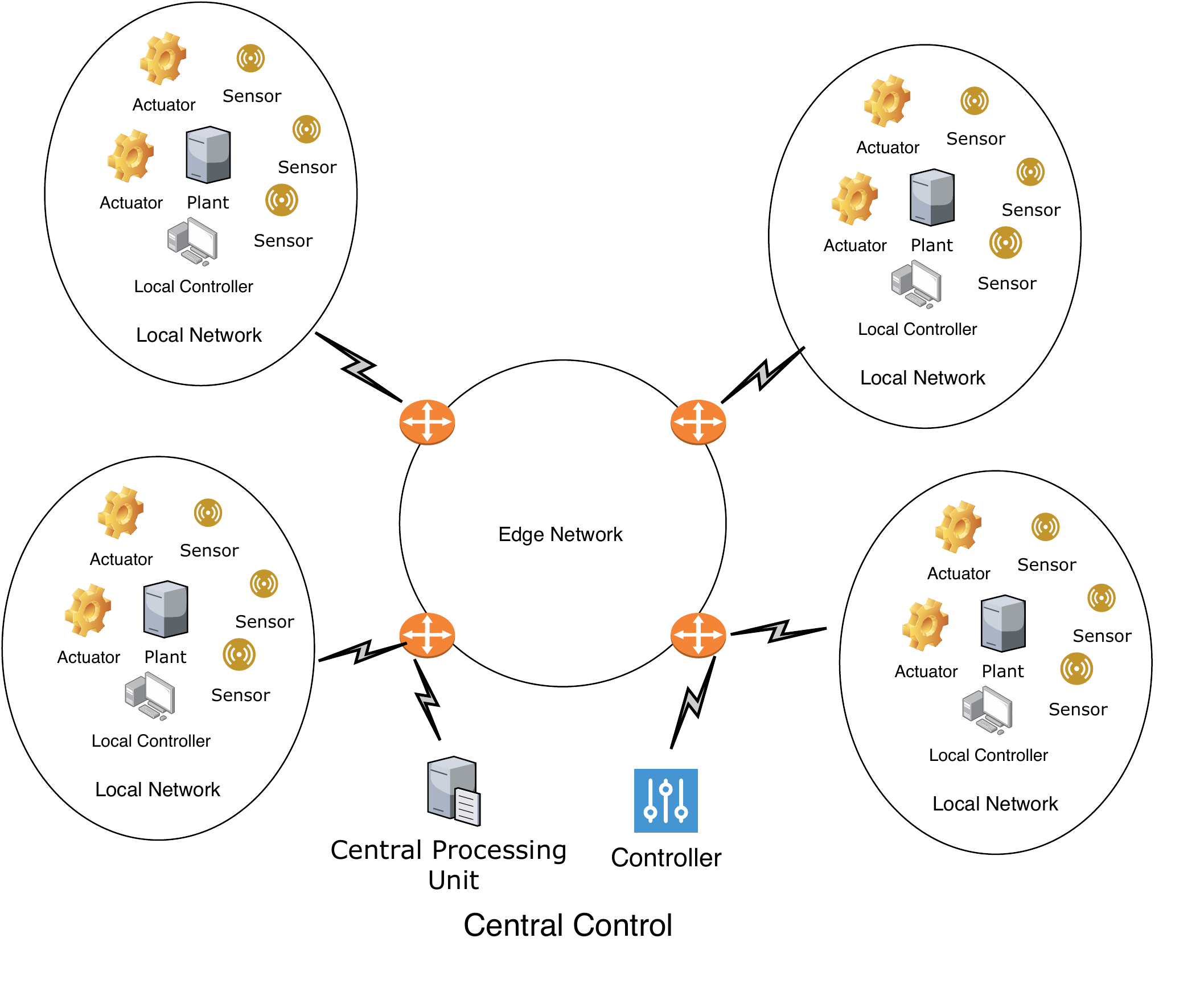}
    \caption{An example of a networked control system.}
    \label{fig:NCS}
\end{figure}

CPSs are the integration of physical processes, networking, and computation. Physical processes are controlled using a feedback loop, where the physical process has an influence on the computation of the system and vice versa \cite{8434710, broo2021cyber,khujamatov2021iot,pivoto2021cyber}. As shown in Fig. \ref{fig:NCS}, components of the NCS, such as sensors, controllers, and actuators, are connected via a communication network, such as an Ethernet-based Fieldbus, a wireless network, or even the Internet. %The network  is used to exchange information between control agents.
An NCS is used to control a physical object or process and, through control feedback, adapts to changing environment conditions in real-time. Traditionally, wired network technology has been used in NCSs due to its reliability and stability. However, there has been a recent shift towards wireless NCSs due to ease of installation, flexibility, including mobility, and lower costs \cite{gupta2009networked}.

In general, communication between agents (, i.e. sensors, actuators, controllers) in an NCS has traditionally been implemented using a time-periodic approach, with agents communicating at regular intervals. One of the difficulties with this approach is determining the sampling period (how frequently agents need to communicate). The sampling period is usually kept low to avoid data loss during system transients. In NCSs, the sampling period is typically set to approximately 20\% of the total available network bandwidth \cite{mazo2008event}. However, allocating a constant sampling period is not always the best approach because it either negatively affects the system performance during transients by selecting too low a sampling period or chooses a high sampling period, which wastes limited network bandwidth by sending redundant control signals during steady-state. To address this issue, event-triggered control techniques have been proposed in which the communication is not time-periodic but the communication frequency increases during transients in the system and decreases during steady-state \cite{guinaldo2016distributed}. Event-Triggered Control (ETC) reduces the continuous utilization of network resources observed in time-periodic communication by acting only when the relevant information is available. These actions include the transmitting of a packet from a sensor to a controller or rescheduling the control tasks when several tasks are running concurrently on the same processor \cite{tiberi2013simple}. During execution intervals, ETC operates in an open-loop mode until the next update arrives. 

In recent years, Machine Learning (ML) techniques have been combined with ETC to improve communication and control performance.  In ETC, event-triggered condition is state dependent and full states information is required by the controller so that it can execute the next update \cite{peng2018survey}. It is not straightforward, however, to linearize the model uncertainties for the ETC \cite{liu2018co} for systems that deal with dynamic changes such as in factory environments or where the system's dynamics change as the load changes, such as in smart grid systems. ML can be used to obtain a model estimate for the controller in order to implement ETC. In conjunction with ETC, ML can be implemented for example in smart grid systems, consensus-based Unmanned Aerial Vehicles (UAVs), and a wide variety of other applications. The only disadvantage of combining ML and ETC is that it will increase the controller's computational load, as ML requires a large amount of data and computations based on that data.
In the literature, ML has been used in combination with ETC to accomplish three fundamental goals:
\begin{itemize}
    
 \item \textbf{Model dynamics learning:} ETC requires access to the system model in order to work effectively, therefore, ML is widely used to learn plant models \cite{solowjow2018event}. Unknown model parameters can be learned or improved by designing a state estimator using ML algorithms. Different ML approaches, such as Statistical Learning (SL), Reinforcement Learning (RL), and Neural Networks (NNs), have been developed to learn and update system models in order to make the model robust against disturbances and uncertainties. By using ML, we can even reduce the requirement for an actual model of the system, as the controller will learn the dynamics of the system using ML\cite{baumann2018deep}.

\item \textbf{Solving an optimization problem:} It is often difficult to solve an optimal control problem for a nonlinear system because the equations are partial differential and usually do not have a closed form solution. For example, solving the Hamilton–Jacobi–Bellman (HJB) equations is a difficult problem in the field of optimal control. ML techniques such as RL or Adaptive Dynamic Programming (ADP) are used to find an approximate solution to optimization problems such as the HJB equations.

\item \textbf{ML for joint learning and optimization:} While RL and ADP have been applied to solve nonlinear optimal control problems, the dynamics of some systems are excessively complex because of their highly nonlinear nature. This leads to a situation where the system model is partially or completely unknown to the controller, and learning the model should be taken into account prior to executing the control optimization step. A combination of RL and NN has been used to simultaneously learn the system dynamics and solve the optimization problem. The Actor-Critic-Identifier (ACI) approach, which is based on RL and is presented to approximate the Hamilton-Jacobi Bellman (HJB) equation, is widely used in this context. Typically, three NN structures are employed in ACI, with actor and critic NNs approximating the optimal control and optimal value functions, respectively, and NN identifier approximating the uncertain system dynamics \cite{bhasin2012actor}. Typically, in ML for joint learning and optimization problems, identifier NN is used to learn system dynamics and critic RL is used to solve event-triggered optimization problems.
\end{itemize}

While numerous techniques have been proposed in the literature to address the aforementioned goals, to the best of our knowledge, no survey has been conducted on how ML can be applied in ETC systems to address changing dynamics, resource management, uncertainties, and disturbances. To address this gap and to stimulate further research and innovation in this area, we present a comprehensive survey of the state-of-the-art of ML-based ETC. We provide a detailed discussion on ML for ETC, divided into three sections: ML for learning model dynamics, ML for optimal control and communication problems, and ML for joint learning and optimization. We have also categorized the state-of-the-art by determining whether ML can be used to learn control behavior (e.g., control inputs), communication behavior (decisions), or both. Both the control and communication policies are identified in the state-of-the-art overview tables. A number of promising research directions and trends are also discussed.

%In this study, we provide a comprehensive review of ML applied in ETC systems to address changing dynamics, resource management, uncertainties, and disturbances. 

In summary, this study makes the following major contributions:

\begin{itemize}
\item{Classification of the ML-based ETC literature: By analyzing ML-based ETC methods presented in \cite{solowjow2018event, solowjow2020event, baumann2018deep, baurnann2019event,yoo2019event,sahoo2015neural, schluter2019event, li2018adaptive, li2022event, chen2021model, liu2021neural, hashimoto2019learning, gao2020event, liang2020neural,wang2022event,  yu2022event,  guo2019event,george2019distributed, zhang2016event, zhang2017event, wang2017event,cui2019event, yang2018event,yang2020event, qin2022event, xue2022event,su2020integral, wang2017improving, zhao2022goal, wang2018learning, yang2019adaptive,huo2022adaptive,zhang2021adaptive, yang2019event, vamvoudakis2018model, yang2018adaptive,bai2021event, wang2019self, sun2022event,lu2022event,yang2017event,ran2022optimizing,wang2017mixed,zhang2021event,huo2021adaptive,xu2021single,liu2022data,xue2021event,li2021event,tan2019event}, ML-based ETC can be classified into three parts depending on the purpose of machine learning use; ML for model dynamics learning, ML for solving a control optimization problem, and ML for joint system dynamics learning and control optimization.}

\item{Classification of ML-based ETC literature in accordance with learning techniques including SL, NN, RL, and Deep Reinforcement Learning (DRL). We also explain the advantages and disadvantages of these learning methods.}

\item{Classification of ML-based ETC literature by control and communication properties, including control structure and controller.}

\item{Reviewing articles that use ML to learn control laws or communication laws, or both, where a learning agent optimizes control actions and communication decision are made.}

\item{Presentation of selected open issues and research trends in ML-based ETC.}

\end{itemize}

The rest of the paper is organized as follows. We provide a brief overview of ETC and ML in Section~\ref{sec:Preliminaries}. In section~\ref{sec:ML_DL}, we present ML techniques used for model dynamics learning in combination with ETC. Optimization-based ML is discussed in section~\ref{sec:ML_OP}. ML articles that discuss joint learning and optimization are reviewed in section ~\ref{sec:ML_JT}. Open issues and future research directions for ML-based ETC are discussed in section~\ref{sec:future_works}. Finally, we conclude our discussion in section~\ref{Sect:Conclusions}.

\section{Preliminaries} \label{sec:Preliminaries}

\subsection{Event-triggered and Self-triggered Control}
Traditional time-triggered networked control approaches use a fixed sampling period, leading to time-periodic communication between the agents in the system. This is often ineffective as it uses network resources even when no updates are required. ETC reduces the continuous utilization of network resources observed in time-periodic communication by acting only when the relevant information is available or needs to be communicated. This makes ETC reactive in nature to when an event is detected. Another, related event-driven control approach, Self-Triggered Control (STC), is proactive and predicts the occurrence of an event based on the system model and current measurements \cite{ijaz2018self}. ETC requires extra hardware resources to continuously monitor the output of the system, which may increase the cost and complexity of the system \cite{wang2021neural}. To overcome this problem, STC was proposed in which the next sampling time is calculated at the current instant and the output of the system is only monitored at sampling instances \cite{heemels2012}.  Numerous studies have attempted to eliminate the need for continuous monitoring in ETC. For instance, continuous communication and self-state monitoring are avoided in\cite{ran2020event}. In \cite{wang2021cooperative}, mixed time- and event-triggered observers are presented to estimate the state of a system with discontinuous monitoring. The most significant benefit of merging time-triggered observers and event-triggered observers in an architectures is that it removes Zeno behavior by default.

Choosing suitable event-triggered threshold is crucial due to determining communication interval and controller updates. Conventionally, a state-dependent static threshold was employed, with communication occurring only when the difference between the current state and the previously transmitted state deviates from a predefined constant threshold \cite{tian2019probabilistic}. Efforts have been made to make this threshold dynamic. In the dynamic event-triggered threshold case, the triggering not only depends on the output of the state but also on the internal variables of system dynamics to adjust the triggering mechanism dynamically with time \cite{ge2020dynamic}. Another triggering threshold mechanism is adaptive event-triggering, in which an optimization algorithm obtains a triggering threshold which is not only dynamically changing but also adaptive to the change in system dynamics \cite{gu2017adaptive}. To avoid Zeno behaviour and guarantee the minimum inter-sampling time, a hybrid event-triggering technique is proposed in \cite{zhao2020hybrid}. While proposed event detection is not continuous, a minimum inter-sampling time is defined such as in periodic communication, which will ensure that the system will avoid Zeno behaviour. Event-detection is performed after this minimum sampling time has elapsed. Some recent advances in ETC are discussed in the following survey paper \cite{peng2018survey, ge2021dynamic, chen2020often}. ETC has been used recently in various fields considering various uncertainties such as delay and packet loss. For example, ETC is applied to asynchronous control of cyber-attacks in \cite{ran2022adaptive}. Actuator non-linearity, and sensor saturation are also considered under a new design of ETC based on fuzzy Markov jump systems in \cite{ran2020event}.

%{\color{blue} The convergence time of multi-agent systems utilizing consensus mechanisms is a trendy topic these days, and adopting event-triggered control to achieve a fixed time of convergence is gaining momentum. This finite-time event-based control technique has been explored in \cite{ran2020event} for developing consensus in second-order multi-agent systems. This work does not require continuous state monitoring or communication with neighbors. Additionally, they take input delays into account by backstepping to ensure that the system's true velocity can match the virtual velocity in finite time. An event-based membership function-dependent asynchronous controller for the interval Type-2 Markov jump system was designed in \cite{ran2020event}. The article considers the uncertainties of the system and how event-triggered control is used to reduce the communication burden.} 

%In developing an event-triggered controller, a model of the plant is required, which is not always available to the controller in practical scenarios because many systems are highly nonlinear and too complex to model, or they have high levels of disturbances, making it difficult to specify a precise model of the system. To overcome the problem of the lack of a dynamic model of the system, the combination of {\color{red}ML} with ETC has been suggested in recent years, in which the controller learns both the control and communication laws without requiring a model of the system. The next subsection provides an overview of the ML approaches used in ETC applications.

\subsection{Machine Learning}
ML is a set of algorithms that make decisions based on available data in order to predict and/or optimize the performance of a system. A good definition of what learning involves is as follows: “A computer program is said to learn from experience E with respect to some class of tasks T and a performance measure P if its performance at tasks in T, as measured by P, improves with experience E”\cite{mitchell1997machine}. In ML-based ETC, machine learning is used to achieve various goals, including learning model dynamics, solving optimal control problems, and joint learning and optimization. The methods used to achieve these goals are statistical learning (SL), neural networks (NN), reinforcement learning (RL), and Deep RL (DRL). These methods are briefly explained in the following. 

\subsubsection{Statistical Learning}
SL approximate solutions to complicated control problems that are costly to solve exactly \cite{koltchinski2000statistical}. To analyze and learn from data sets, SL focuses on statistical properties. For example, Empirical Risk Minimization (ERM) is a concept in SL that defines algorithms that yield theoretical bounds on performance \cite{koltchinski2000statistical}. In general, a SL algorithm is fed a training set as an input, which is sampled as an unknown distribution and labeled with a target function, and the output is a predictor that finds the minimized error with respect to an unknown distribution and target function. Since the learner does not have any information about the unknown distribution and target function, the true error is not accessible directly to the learner. An error that can be calculated by the learner is the training error that a classifier incurs over the training sample, which is also known as an empirical error or empirical risk. ERM searches for the solution that minimizes the empirical error. SL has been used in existing ML-based ETC works to learn the dynamics of the model. For example, SL is used to learn a new model based on statistical properties of inter-communication time. It is also used in combination with controllers such as Linear Quadratic Control (LQR) or Model Predictive Control (MPC) and makes these controllers robust against uncertainties via increasing prediction and estimation.
As SL is a combination of statistics and ML, building an SL model requires a good understanding of statistical properties of the data.

\subsubsection{Neural Networks}
A neural network (NN) is an ML technique in which data is processed similarly to the way the human brain works with sensory data. When an input is provided to the NN, it will generate the best possible results via adopting to changing input without redesigning the output criteria. For example, Radial Basis Function (RBF) NNs have been used as a tool for modeling nonlinear functions in control engineering due to their simple structure and good accuracy~\cite{papadimitrakis2022active}.
RBF networks can approximate an unknown function with a linear combination of a group of nonlinear functions, called base functions. In nonlinear systems, the system dynamics are unknown, which means the ETC framework cannot be directly applied.  RBF NN is a powerful method applied in many areas of engineering due to its flexibility in adapting data distribution, fast training, and short run-time\cite{golnaraghi2020predicting}. However, NN usually requires a lot of data in comparison with more traditional ML algorithms, which makes the approach not suitable for many problems where data is limited.
%{\color{red} As a result, RBF NN approximation is used to approximate unknown continuous functions \cite{li2018adaptive}. RBF NN is different from regular NN in that it represents input data as a statistical distribution (usually Gaussian). The main advantage of RBF NN over artificial NN is that it has a simple topological structure. Because of this, RBF NN needs less computational time for both training and testing. \cite{golnaraghi2020predicting}.}

\subsubsection{Reinforcement Learning}
RL is an ML method in which agents take actions by trial and error and, in return, receive rewards based on those actions from the environment in which they operate. At each time step, the agent takes an action that may result in a transition to a new state of the environment. Then, the agent receives a reward based on the quality of this transition. RL agents estimate policy and value functions. The value function looks at the agent's current situation in the environment, while the policy function looks at how the agent makes decisions.

 RL can be categorized into three types: actor-only, critic-only, and actor-critic methods, where the terms "actor" and "critics" are used instead of policy and value function, respectively \cite{singh2021reinforcement}. Although the value function method has been successful with discrete lookup table parameterization, this method failed to generalize when applied to continuous function approximation. Q-learning and deep Q-learning are examples of this method. On the other hand, policy function methods have strong convergence guarantees in comparison with the value function method, which is quite inefficient even when applied to simple examples with few states\cite{peters2005natural}. While the policy function approach has been successful in continuous and stochastic environments and has a faster convergence, value functions are more sample efficient and steady. Therefore, actor-critic methods merge these two approaches to benefit from both and achieve a better result. In the actor-critic approach, the actor performs an action on the environment, and the critic evaluates the values of the action and sends feedback information to the actor \cite{yang2017event}. Based on our review, RL algorithms using both critic and actor-critc methods have been developed to learn the model of the system, solve an optimization problem, and perform joint learning and optimization. However, the actor-only method is not used in any of the reviewed articles. Using RL means taking action based on rewards helps to learn the dynamics of a system accurately. However, RL is not preferable for solving simple problems or for solving problems that need a lot of data.

\subsubsection{Deep Reinforcement Learning}
Traditional ML approaches exhibit problems when dealing with high-dimensional data, which has recently become more widely available. This has led to the development of the concept of deep-learning (DL)\cite{nasir2021review}. DL is a subset of ML based on NNs that uses multiple layers of non-linear information processing for both supervised and unsupervised feature extraction of data.\cite{nguyen2019machine}. DL can be combined with RL, which helps to overcome RL's limitation to domains with fully observed and low-dimensional state-spaces. This combination, deep DRL, can easily find compact low-dimensional features in high-dimensional data \cite{ashraf2021state}.

\section{Learning Model Dynamics}\label{sec:ML_DL}
In this section, we review works that have used ML to learn the dynamics of the system model reported in Table. \ref{tab:comp1}. As the accuracy of the available system model has a direct impact on control performance, it is possible to improve closed-loop performance by updating and improving the model during operation using data. The available model can be improved by learning an uncertainty compensation model and designing a state estimator supported by ML or by learning unknown model parameters. In ETC, the efficacy of the control approach dependents on the availability of an accurate dynamic model. ML can be used to learn such a model. Combining ML with ETC makes the control system more robust against disturbances and uncertainties, while computational load as well as communication bandwidth can be reduced significantly. 

\subsection{Learning Model Dynamics with Statistical Learning}
Although learning methods have the potential to improve system performance, performing a learning task is expensive (e.g., including communication resources and computation costs). Therefore, several articles \cite{solowjow2018event, solowjow2020event, baurnann2019event} consider event-triggering rules for model learning, which determine when a new model should be learned based on statistical properties of intercommunication time. These articles develop learning triggers through the derivation of model-induced probability distributions and the observation of intercommunication times. Additionally, statistical estimates can be obtained by using concentration inequalities such as Hoeffding’s inequality and the Dvoretzky-Kiefer-Wolfowitz (DKW) inequality.

A novel Event-Triggered Learning (ETL) approach is applied to linear Gaussian systems by combining State Estimation and SL in \cite{solowjow2018event}. This combination will result in higher prediction accuracy and cheaper communication costs, even when compared to Event-Triggered State Estimation (ETSE), because the model will be improved through learning. ETSE's effectiveness in reducing communication is entirely dependent on the prediction's accuracy, or in other words, the model's quality. When the present model's prediction performance is low, learning experiments are triggered to improve the model using the available data. Model learning can be solved with a standard least-square estimator. Hoeffding’s inequality is considered here as the concentration inequality to quantify the confidence level. The approach demonstrated a reduction in communication effort in both simulation and hardware implementation of a cart-pole system. The same authors' subsequent work \cite{solowjow2020event} extended ETL by adding a Kalman filter and demonstrating the impacts with new illustrative use cases. Moreover, they used the DKW inequality in addition to Hoeffding’s inequality to provide more detailed statistical information because it provides bounds on the empirical Cumulative Distribution Function (CDF). However, no disturbance is considered in \cite{solowjow2018event} and \cite{solowjow2020event} and the results are only developed for linear Gaussian systems. The work could be extended to non-linear dynamic systems to investigate how ML may support ETL with non-linear systems.

\begin{figure} [h]
    \centering
   \includegraphics[width=\linewidth]{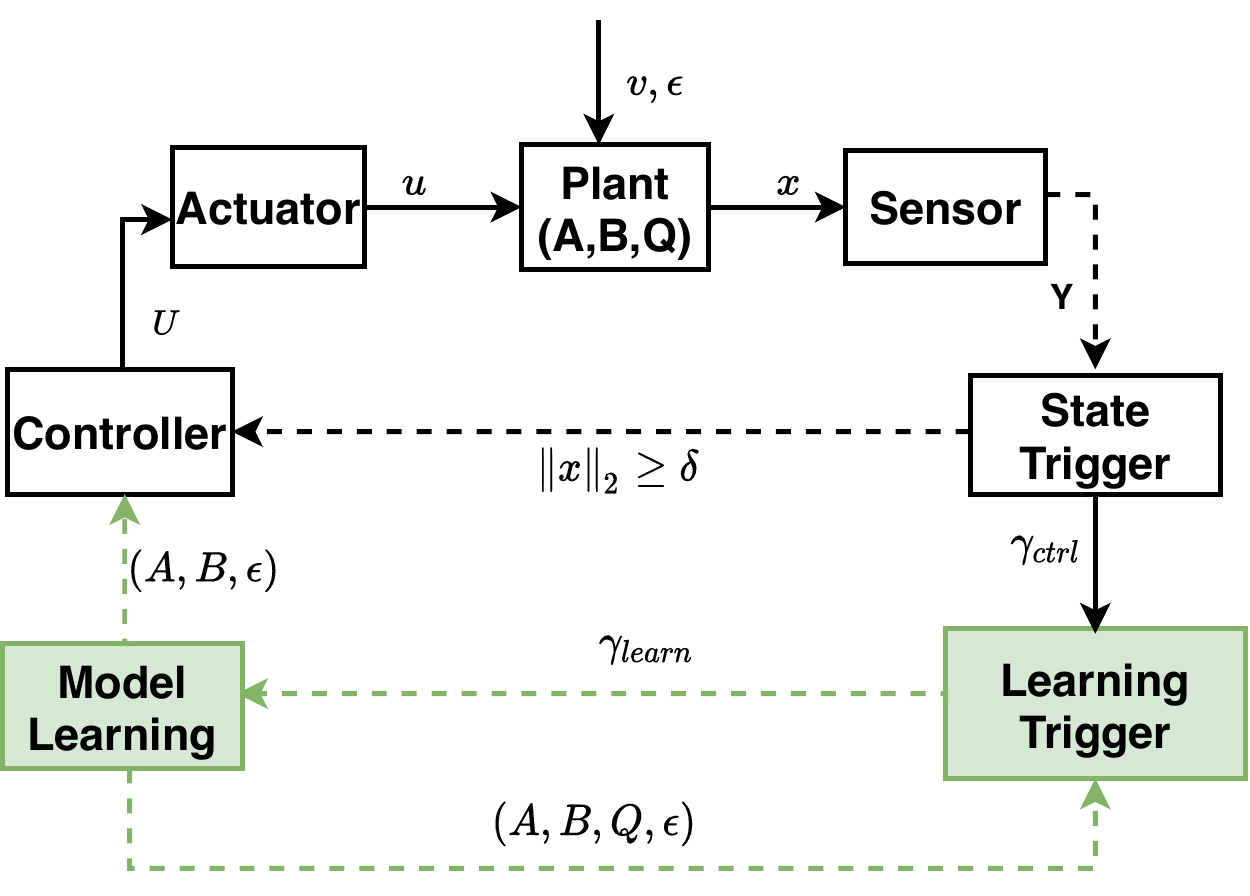}
    \caption{Event triggered learning diagram\cite{baurnann2019event}}
    \label{fig:ETL}
\end{figure}

The framework used in \cite{solowjow2018event} is further developed in \cite{baurnann2019event} to include a control loop based on the concept of event-triggered pulse control mixed with SL to learn dynamic models, as illustrated in Fig. \ref{fig:ETL}. This strategy is beneficial when the initial model is poor or when the dynamics have changed, causing learning to occur. A learning trigger decides whether or not the system model is accurate enough. If the accuracy is insufficient, the learning of a new model is triggered (see the green section in Fig. \ref{fig:ETL}). The authors introduced two different triggers, (i) state trigger ($\gamma_{ctrl}$), which initiates communication of control commands when $\delta$ (a user-defined threshold) is exceeded; and  (ii) a learning trigger ($\gamma_{learn}$), which initiates learning in the case of poor performance. When an event occurs, the system is reset to its equilibrium state via the application of a pulse whose duration is determined by the dynamics of the system. A plant is considered with sensors and actuators with noise and disturbance ($v$ and $\epsilon$ ). Hoeffding’s inequality is used to quantify the confidence level in the estimation. According to the authors, two major characteristics of this method are its ability to adapt to changing dynamics and to discover an appropriate alternative for the integral control component used in periodic control. Numerical simulation demonstrates that learning system dynamics has the effect of reducing communication effort, coping with load disturbances, and changing dynamics. All numerical simulations, however, were conducted on first-order systems, not higher-order systems.

In contrast to \cite{solowjow2018event}, \cite{solowjow2020event}, and \cite{baurnann2019event}, which use the model of the system for prediction of the next trigger event, which is based on communication, the authors of \cite{schluter2019event} use the model for control purposes, and triggering is based on control performance. LQR is used in \cite{schluter2019event} to minimize the expected value of the cost function by combining it with SL theory. Model learning is activated when performance improvements are required. Hoeffding’s inequality and the Chernoff bound are used as concentration inequalities to obtain an effective trigger with enough theoretical guarantees. In this approach, when the empirical cost of the Riccati equations over a finite horizon exceeds the Chernoff Bound, learning is triggered. Least square estimation is used to perform the learning. The proposed method is implemented to control the pole-balancing performance of a rotary pendulum. The authors validate whether the trigger is capable of detecting these changes and evaluate model accuracy by adjusting the ball joint and magnetic weights in the pendulum. SIMULINK is used to implement a switched controller with a sample rate of 500 Hz, and an LQR with an integrator is used to stabilize the upright position. Least squares estimation is used to perform the learning.

While the previously described works applied event triggering to the learning process to build an accurate model, the authors of \cite{yoo2019event} study event triggering applied to the control process. Here, event-triggered MPC is combined with ERM, which makes the control system adaptive and robust to uncertainties and state errors. MPC is a form of optimal control that can tackle multi-variable systems and handle hard constraints on input, state, and output variables by solving a finite-horizon open-loop optimization problem \cite{schwenzer2021review}. The main goal of \cite{yoo2019event} is to attenuate the unknown disturbance by designing an uncertainty compensator using ML, or, more precisely, by using ERM with kernel regression to predict the system state subject to uncertainties and learning an uncertainty compensation model to obtain the bound of uncertainty. In fact, by applying the ERM as an SL approach, restrictions that require a known upper bound of uncertainty or a known structure of uncertainty (constant or harmonic), which are standard assumptions when designing robust MPC or adaptive control, are eased. It is worth noting that \cite{yoo2019event} does not use online ML and the compensator is not updated during the control operation.

\subsection{Neural Networks (NNs)}

\begin{figure} [h]
    \centering
   \includegraphics[width=\linewidth]{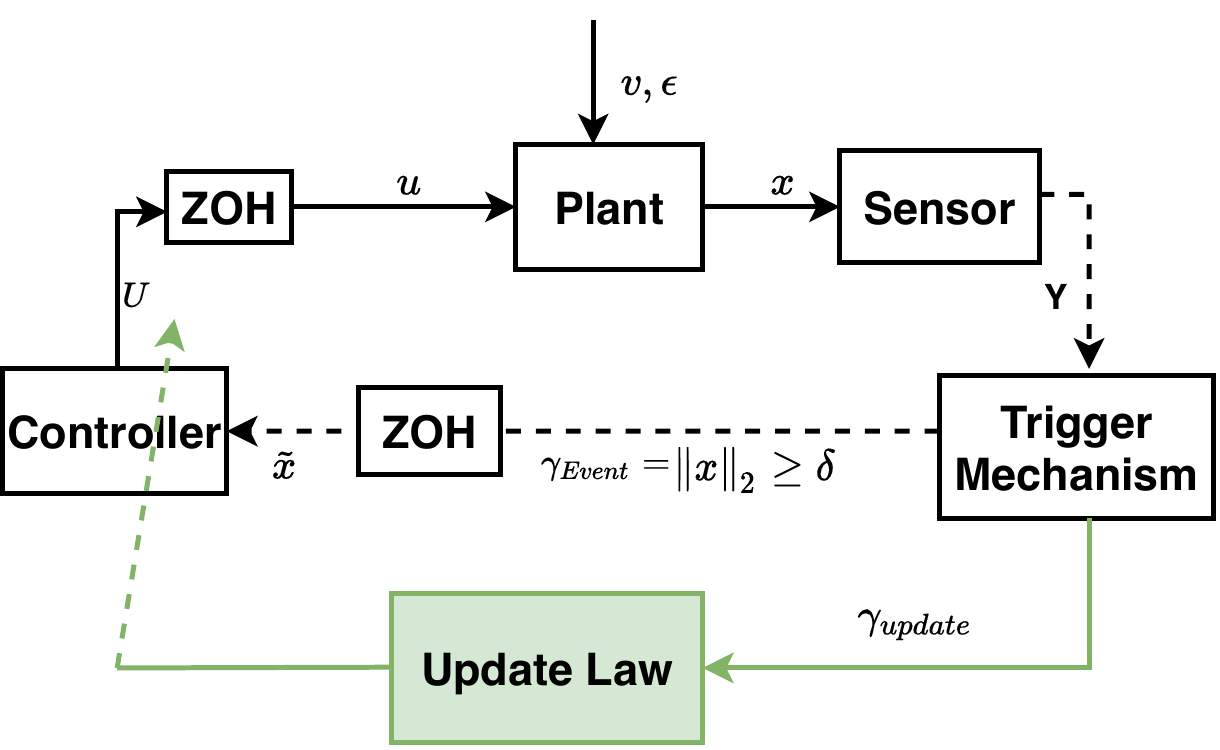}
    \caption{NN-based ETC\cite{sahoo2015neural}.}
    \label{fig:NN}
\end{figure}

An online approximate ETC is designed for nonlinear multi-input-multi-output (MIMO) uncertain systems using NNs in \cite{sahoo2015neural}. The controller is approximated by utilizing a linearly parameterized NN. A system state vector-based event-triggered condition is described, and the condition is made adaptive (state dependent, monotonically increasing) in order to achieve a trade-off between ETC approximation and resource utilization. A novel NN weight update law, as shown in Fig. \ref{fig:NN}, ensures the reduction in network resource utilization and relaxes the required knowledge of the whole dynamic system. NN weights are updated through an event trigger mechanism in an aperiodic manner, as illustrated in Fig. \ref{fig:NN}. As a result, the proposed method requires less computation than traditional NNs that update periodically. Event-triggered communication in \cite{sahoo2015neural} is also a function of NN weight estimates and system states, whereas traditional ETC is a function of system states, resulting in overall less computation. To demonstrate Lyapunov stability, the event-triggered system is modeled as a nonlinear impulsive dynamical system. According to the authors' simulations, the strategy produced a 45 percent reduction in computation burden when compared to the periodic method.

In contrast to \cite{sahoo2015neural}, which uses a linearly parameterized NN to approximate the controller,\cite{li2018adaptive,gao2020event,liu2021neural} use RBF NN to approximate unknown functions.
Similar to \cite{sahoo2015neural}, the NN weights are also updated at event-triggered instances in \cite{li2018adaptive}. An adaptive ETC is proposed without making any assumptions about input-to-state stability, even while the model of the plant is uncertain, in order to formulate a practical dynamic system. The adaptive ETC approach is used in conjunction with a NN to develop a pure feedback controller for nonlinear systems with nonlinear model uncertainty. The proposed method uses adjustable dynamic sampling states to update the controller model and the adaptive control law. In contrast to a constant state-dependent threshold, a dynamic threshold is employed for ETC. The NN weights and controller are only updated when the desired control specification cannot be guaranteed. While utilizing Lyapunov stability to evaluate the model, the proposed ETC reduces computation and communication load. Since general state-feedback systems are not considered in \cite{li2018adaptive}, an event-triggered based controller for pure feedback systems is proposed using adaptive NN tracking in \cite{gao2020event}. Here, the mean value theorem is used to transform the pure-feedback nonlinear system into a strict-feedback nonlinear system. A NN approximates the output tracking error in finite time to bound the error close to zero using finite-time prescribed performance, which will guarantee the same performance for both transient and steady-state. ETC is used to obtain a large (both fixed and variable) threshold. The adaptive NN ensures that all signals in a closed loop are bounded, which is verified by applying the Lyapunov stability theory. 

\cite{li2018adaptive,gao2020event} an adaptive ETC problem is studied for a class of electromagnetic suspension systems with unknown parameters utilizing back-stepping technology in \cite{liu2021neural}. RBF NN is used to approximate the unknown functions. An ETC with a fixed threshold strategy and the relative threshold control approach are devised and compared in order to reduce communication resources. The authors proposed for future work the development of an intelligent controller that can switch between fixed threshold and relative threshold based on threshold size.

 An adaptive ETC problem is extended for non-affine nonlinear multiagent systems with uncertainties including dynamic disturbance, model-free dynamics, and dead-zone input in \cite{liang2020neural}. In \cite{liang2020neural}, RBF NN is used to approximate the unknown function, similar to \cite{li2018adaptive,gao2020event,liu2021neural} and the ETC back stepping design procedure is combined. Both unknown dead-zone and model-free dynamics are considered simultaneously in \cite{liang2020neural} for multiagent systems. Adaptive ETC RBFNN back stepping control is studied more, including in \cite{wang2022event} for MIMO switched nonlinear systems with output and state constraints and non-input-to-state practically stable (ISpS) model-free dynamics, in \cite{li2022event} for completely unknown nonlinear functions with dynamic gain, and in \cite{yu2022event} for under-actuated marine surface vessels using an NN-based disturbance estimator. Adaptive ETC RBF NN is also presented in \cite{chen2021model} for a class of single-input-single-output uncertain nonlinear continuous-time (CT) systems by integrating input-to-state linearization techniques, impulsive dynamical system and RBF NN with adaptive ETC threshold.

\subsection{Reinforcement Learning}

%%%%it should be comment I did not see any STC!
%A joint learning algorithm is proposed in \cite{hewing2018cautious} based on RL and STC. The authors use a Gaussian process (GP) formulation model to learn the dynamics of the plant, as obtaining a model of the plant is difficult due to the system's dynamics being too complex or highly non-linear. An MPC is formulated, and the optimal control problem is solved for each time step based on the dynamics learned by the GP. 

An STC based on Gaussian Process (GP) regression %is introduced in \cite{hewing2018cautious}  and 
is developed in \cite{hashimoto2019learning} for NCS with unknown system dynamics. In \cite{hashimoto2019learning}, a joint learning algorithm is proposed that uses RL to learn the dynamics of the plant and a self-triggered controller to reduce the number of communication time steps for NCS. An infinite horizon optimal control problem has been formulated that takes into account both the control and communication costs. The MPC problem is solved by using the GP dynamics of the plant to obtain control input for STC. The authors divide this framework into two parts, the execution phase and the learning phase. During the first phase, the STC is implemented in an epsilon-greedy \cite{dabney2020temporally} fashion, in which a random control input with one-step intercommunication time is sampled; otherwise, the computed optimal control and communication policy is executed. In the second (learning) phase, the learning agent uses the training data to update the GP model of the plant and compute the optimal control and communication policies.

Deep RL (DRL) is based on combining RL with a deep learning algorithm to generate an efficient learning algorithm. In \cite{baumann2018deep}, DRL has been used to simultaneously learn control and communication behaviour of a model-free system and then utilise this DRL for ETC to reduce sampling. A RL problem is formulated as a resource-aware control strategy, where the learning agent optimizes its control input and communication decisions to maximize the expected reward over the time horizon. The reward function comprises two terms, one is to capture control performance, and the other gives a reward for time steps without communication. Two learning ETC approaches are proposed. In the first learning approach, only communication is considered using feedback control, but in the second learning approach, both control and communication are simultaneously considered, which is called end-to-end learning. In terms of ETC, end-to-end learning emphasizes learning of both communication and control models simultaneously, rather than separating them. During the training of the agent, it receives negative rewards for bad performance (early termination of the episode) and for every communication. In an RL task, the agent’s interaction with the environment is divided into episodes. An agent receives a constant positive reward to prevent an unwanted early termination of the episode. The authors use joint learning for control and communication, which reduces communication by using a parameterized action space Markov decision process. 

\subsection{Actor-critic RL}
Adaptive tracking control based on dead-zone event-triggered RL is presented in \cite{guo2019event} for a nonlinear CT system with external disturbances and unknown dynamics without the Persistently Exciting (PE) condition and initial stabilizing control. To approximate an unknown long-term performance index, controller, critic, and action NNs are used. To demonstrate the developed controller's performance, an autonomous underwater vehicle model was chosen for simulation. The ETC threshold increases monotonically, and the system is Uniformly Ultimately Bounded (UUB).

\begin{table*}[h]
\caption{Existing ETC systems proposed in the literature - ML for dynamic model learning}  \label{tab:comp1}
\scriptsize
\centering          % used for centering table
%\begin{tabular}{| P{0.8cm} | P{0.8cm} | P{.6cm} |P{.8cm} |P{.5cm} |P{.7cm} |P{.8cm} |P{1.5cm} |P{0.8cm} |P{.7cm} |} 
\begin{tabular}{| c | P{1.2cm} | P{1.2cm} | P{1.4cm} |P{.8cm} |P{.7cm} |P{1.4cm} |P{1.5cm} |P{1.4cm} |P{.8cm} |} 
%\begin{tabular}{| c | c | c |c |c |c |c |c |c|c|}    % centered columns (4 columns)
\hline
 %inserts double horizontal lines
Reference & System Type & Learning Approach & Communication policy & Control policy  & Multi-agent & \seqsplit{Centralized~/~Decentralized~/~Distributed}  & Control Mechanism & \seqsplit{Experimental~Validation}  \\[0.5ex]  % inserts table
%heading
\hline
\cite{solowjow2018event} , \cite{solowjow2020event}& Linear Gaussian System & SL & \xmark & \cmark & \xmark & \ Cent. & State Feedback Controller & \cmark \\
\hline
\cite{baurnann2019event}& Linear time-invariant & SL & \xmark & \cmark & \xmark &  \ Cent. & Simple Gain & \xmark\\
\hline
%\cite{beuchert2019hierarchical}& Nonlinear (DT) & SL & \xmark & \cmark & \xmark &  \ Cent. & -  & \cmark\\
%\hline
\cite{yoo2019event} & Nonlinear (DT) & SL & \xmark & \cmark & \xmark &  \ Cent.  & MPC & \xmark\\
\hline
\cite{sahoo2015neural} &\ Affine Nonlinear (CT) & NNs  & \xmark & \cmark & \xmark & Cent. &  State Feedback Controller & \xmark \\
\hline 
\cite{schluter2019event}& Nonlinear & SL & \xmark & \cmark & \xmark & \ Cent. & NN  & \cmark  \\
\hline 
\cite{li2018adaptive}&  Non-affine Nonlinear & RBFNN & \xmark & \cmark & \xmark & Cent. & Backstepping Control & \cmark \\
\hline 
\cite{liu2021neural} & Nonlinear & RBFNN  & \xmark & \cmark & \xmark & Cent. & Backstepping Control & \xmark \\
\hline 
\cite{liang2020neural} & Non-affine Nonlinear & RBFNN & \xmark & \cmark & \cmark & Distributed & Backstepping Control & \xmark \\
\hline 
\cite{wang2022event} & Nonlinear & RBFNN  & \xmark & \cmark & \xmark & Cent. & Backstepping Control & \xmark \\
\hline 
\cite{yu2022event} & Nonlinear CT & RBFNN  & \xmark & \cmark & \xmark & Cent. & Feedback Linearization & \xmark \\
\hline 
\cite{chen2021model} & Nonlinear & RBFNN  & \xmark & \cmark & \xmark & Cent. & Backstepping Control & \xmark \\
\hline 
\cite{li2022event} & Nonlinear & RBFNN  & \xmark & \cmark & \xmark & Cent. & Backstepping Control & \xmark \\
\hline 
\cite{gao2020event}&  Non-affine Nonlinear &  RBFNN & \xmark & \cmark & \xmark & Cent. &  Backstepping Control & \cmark \\
\hline 
\cite{hashimoto2019learning}& Nonlinear & RL & \cmark & \cmark & \xmark & Dist. & Optimal control  & \xmark \\
\hline 
\cite{baumann2018deep}& Nonlinear& DRL & \cmark & \cmark & \xmark & Cent. & State Feedback Control & \cmark  \\
\hline 
\cite{guo2019event}& Nonlinear (CT) & RL & \xmark & \cmark & \xmark & Cent. & Adaptive Control & \xmark\\
\hline
\end{tabular}
\end{table*}

\section{ML for Optimal (Control and Communication) Performance} \label{sec:ML_OP}

In this section, we discuss papers that employ ML to address optimization problems, as reported in Table. \ref{tab:comp2}. RL and Adaptive Dynamic Programming (ADP) can satisfy both optimal control policy and optimal performance simultaneously\cite{yang2019event}. In general, the RL method includes an actor to improve performance via interacting with the external environment and a critic to evaluate the control performance of the actor\cite{cui2019event}. RL approaches have been applied to solve a variety of optimization problems, including optimal regulation problems \cite{li2021distributed}, robust control problems \cite{mu2019learning}, and differential games, including zero-sum
games\cite{chen2022novel} and non zero-sum games \cite{su2021event}. Moreover, ADP and RL methods were developed to estimate the solution of the Hamilton–Jacobi–Bellman (HJB) \cite{wen2021optimized} and the Hamilton–
Jacobi–Isaacs (HJI) \cite{zhang2021observer} equations. ADP, as a potential technique for obtaining satisfying solutions to HJB equations, can be classified into three primary categories: Heuristic Dynamic Programming (HDP), Dual
Heuristic Dynamic Programming (DHDP), and Globalized Dual Heuristic Dynamic Programming (GDHP)\cite{wang2019self}. 

Recently, ETC has been integrated with RL and ADP algorithms to increase computing efficiency and conserve communication resources. To solve non-convex optimization problems, a distributed stochastic gradient descent algorithm combined with an event-triggered communication mechanism has been proposed in \cite{george2019distributed}. In the following, we have classified papers into two categories based on their approach to solving optimization problems: critic-only method and actor-critic method. In some works, the critic-only method replaces the common actor-critic structure to simplify the iterative framework and implementation process.

\subsection{Reinforcement Learning - Critic-only}

Actor Critic Learning (ACL) has been successfully applied to a variety of robust control problems as a technique that combines dynamic programming and NN to create a highly effective method for solving specific optimization problems \cite{yang2020event}. However, the control system required for ACL implementation must be persistently excited. Concurrent learning (CL) or Experience Replay (ER), which combine historical and current state data, may allow the Persistent Excitation (PE) condition to be relaxed. CL's main idea is to apply batch-like dynamics to parameter estimation dynamics by utilizing recorded input and output data \cite{parikh2019integral}. For instance, \cite{zhang2016event} and \cite{yang2019adaptive} applied CL to guarantee parameter convergence without requiring PE.

Event-triggered Concurrent Learning (ETCL) is presented in \cite{zhang2016event} for solving the HJI equation of a $ H_{\infty} $ control problem for a class of CT nonlinear systems with external perturbation. The authors defined the $ H_{\infty} $ control problem as a two-player zero-sum game in which the control minimizes the cost function in the worst-case disturbance. An adaptive triggering condition is also obtained for the closed-loop system using an ETC policy and a time-triggered disturbance policy. In the ETCL algorithm, a single critic NN is used for implementation purposes. Additionally, a novel critic tuning law based on the CL technique is used, which allows the traditional PE condition to be relaxed. The results are compared with a concurrent RL algorithm used in \cite{yasini2015online} and a synchronous
policy iteration algorithm (SPIA) in\cite{vamvoudakis2012online}, and the ETCL method is found to be superior in terms of performance against disturbances. Notably, concurrent learning-based ETC requires robust estimation techniques because it requires knowledge of state derivatives, which are typically not directly sensed.

An event-triggered optimal control problem with Integral Reinforcement Learning (IRL) is proposed to solve the HJB equation of CT nonlinear systems with partially unknown dynamics in\cite{zhang2017event}. IRL is a class of RL methods, which are developed based on policy iteration and value iteration, using iterative methods to achieve the optimal solution asymptotically by minimizing the integral temporal difference error at each step \cite{peng2021online}. In comparison to \cite{yang2017event}, which uses joint model learning and optimization, this method does not require any NN-identifier to identify the unknown internal dynamics. A single-critic NN is used in \cite{zhang2017event} to approximate the optimal value function and the optimal control policy for implementation. The UUB of critic weights are validated via the Lyapunov theory. However, no disturbance is considered, and the method is dependent on the initial admissible policy. The number of controller updates is significantly reduced during the simulation result learning process.

The method's applicability is limited by the fact that the approaches in \cite{zhang2016event} and \cite{zhang2017event} are dependent on an initial stabilizing control policy and consider an undiscounted cost function. Therefore, the goal of \cite{wang2017event} and  \cite{cui2019event} is to benefit from ML to obtain the event-triggered nonlinear discounted optimal control law that is independent of the initial condition.
 
In \cite{wang2017event}, training NNs using a learning rule resulted in a near-optimal discounted event-based control law that is independent of the initial condition in an adaptive critic framework. Discounted optimal control considers stage costs, which are weighted by a time-varying decaying term \cite{granzotto2020finite}. The discount factor in the cost function can adjust the convergence speed of the regulation design and reduce the final value of the optimal cost function. In \cite{wang2017event}, the mentioned method is applied to industrial systems such as power systems, as an example of an affine nonlinear system. The stability of a closed-loop system is considered an impulsive model, and its stability is determined using the Lyapunov technique. The controller's performance demonstrates that when the discount factor is increased, the optimal cost decreases, validating the results of event-based near-optimal control performance with discounted cost functions. Additionally, controller updates are reduced by up to 66.76\% during the learning process for power applications. However, the proposed method requires knowledge of the dynamic model and constrained control inputs are not taken into account.

Event-triggered $ H_{\infty} $ tracking control is combined with RL for a CT nonlinear system with external disturbances in \cite{cui2019event}. An event-triggered tracking HJI equation is developed based on an augmented system with the tracking error dynamics and a discounted cost function to solve the $ H_{\infty} $ problem. The HJI nonlinear partial difference equation is solved using a novel RL with a critic network that approximates the optimal cost function independently of the initial admissible control policy. The Lyapunov theory is used to determine the stability of closed-loop systems. The proposed method has several characteristics, including UUB of weights in critic NNs and asymptotic convergence of tracking error to zero. The simulation results indicate that ETC requires 55 samples, whereas a time-triggered controller requires 200 samples, demonstrating the reduction in computing burden while achieving asymptotic tracking. Constrained control inputs, on the other hand, are not considered.

Input constraints such as actuator saturation are significant physical characteristics of actuators in industrial applications and must be considered. Therefore, to address the weakness of not considering constrained control inputs in \cite{wang2017event}, input constraints are taken into account in \cite{yang2020event} and \cite{yang2018event} through the use of a discounted cost function. Moreover, while the disturbance policy is updated using a time-driven strategy and the control policy is updated using an event-triggered mechanism in \cite{zhang2016event} and \cite{cui2019event}, both the control and the disturbance policies in \cite{yang2020event} and \cite{yang2018event} are updated using an event-driven mechanism, significantly reducing the computational load in comparison to other works in the literature that only update the control policy in the event-driven mechanism.

ETC is used in conjunction with adaptive critic design in \cite{yang2018event} to study nonlinear systems with mismatched perturbations and input constraints. By defining an infinite-horizon cost function, the robust stabilization problem is transformed into a constrained $H_2$ optimal control problem. As a result of solving the event-triggered HJB equation, the system states are UUB. A single network adaptive critic design, which is used for solving HJB, is tuned via the gradient descent method. All signals in the closed-loop system are proven to be UUB via the Lyapunov method. The proposed method has a limit when applied to nonlinear, complicated systems due to the difficulty of computing the Moore–Penrose pseudo-inverse of the control matrix function.

An event-driven HJI equation associated with a two-person zero-sum game is proposed in \cite{yang2020event} for CT nonlinear systems with a disturbance. An $ H_{\infty} $ control problem with asymmetric input constraints has been proposed. The $ H_{\infty} $ control problem is converted into a zero-sum game that can be solved using ACL. ADP, ACL, and RL algorithms are often similar as they have the same characteristics. ACL uses historical data and instantaneous state data to update both control and disturbance in the event-driven mechanism. Zeno behavior is also excluded without the requirement of properly selecting disturbance attenuation. Then, using ACL, the event-driven HJI equation is solved and its weight parameters are tuned by applying the gradient descent method. UUB is guaranteed using the Lyapunov approach. 
The results indicate that when ACL is used with both event-driven control and event-driven disturbance, the computational load is reduced by up to 60\%. This method is also applicable to systems that have an equilibrium point at the origin. This is a limitation of the  method, as obtaining information about controlled systems and knowledge of their control matrix is difficult in real-world applications. Similar to \cite{yang2020event}, a zero-sum game problem is solved in \cite{qin2022event}, for nonlinear safety-critical systems with safety constraints and input saturation using a barrier function. A critic NN is developed to approximate the optimal safety value function of the HJI equation, and a novel event-triggered scheme is used to obtain the update instant of the control law and the disturbance law. The CL is also used to relax the PE condition.

While \cite{yang2020event,qin2022event} present zero-sum games,\cite{xue2022event, su2020integral} applies an event-triggered IRL algorithm to a non-zero-sum game problem. To address asymmetric input saturation, novel non-quadratic value functions with a discount factor are used in \cite{xue2022event}. To alleviate the need for a comprehensive understanding of the game, an IRL-based coupled Hamilton–Jacobi equation is derived. To relax the PE condition, the weights of a single critic NN are tuned based on the ER method.

While \cite{cui2019event,zhang2016event,wang2017improving,yang2018event,yang2019event,zhang2017event,yang2019adaptive,wang2017event,yang2020event,xue2022event} studied optimal control problems for nonlinear CT systems, \cite{zhao2022goal} investigates ETC near-optimal problems for input-constrained nonlinear discrete time systems with the input-to-state stability (ISS) attribute subject to actuator saturation. First, the robust control problem of the uncertain system is converted to a near-optimal control problem via the designed cost function. Then adaptive ETC is designed to save computational resources. To improve the control performance, a goal representation adaptive critic design is presented, which consists of two NNs, namely the goal network and the critic network. A goal network is used to learn the external reinforcement reward and provide a more efficient internal reinforcement reward for the critic network with non-periodic weight updating.

In \cite{wang2018learning}, an adaptive self-learning control approach is designed with matched uncertainties (for a plant whose model is uncertain) using an event-triggered critic cost control approach. ADP is used to solve the optimal control problem in a learning-based, forward-in-time approximation fashion. An event-triggered cost control approach using a self-learning technique for nonlinear systems is designed. The controller design is transferred to an optimal control problem with an event-based strategy to have a robust optimal control design. The event-triggered threshold dynamically changes in response to changes in the system's states. A NN is used to implement event-based optimal control with stability guarantees using Lyapunov stability. Learning and guaranteed cost control of the proposed method are limited to nonlinear systems with matched uncertainties and do not include unmatched uncertainties. The proposed method could be improved to track a trajectory based on learning.

ML-based ETC for the decentralized structure of nonlinear systems with uncertain interconnections is discussed in \cite{yang2019adaptive}. Decentralized ETC is developed with ACL and ER for a class of CT nonlinear systems with uncertain interconnections. A critic network is used to solve the event-triggered HJB equations related to optimal ETC laws of the subsystems. Gradient descent and ER are used to update the critic network's weights. ER helps to relax the PE condition. The estimated weight vectors used in the critic networks are proven to be UUB through a classic Lyapunov approach. Overall stability is also achieved based on the stability of decentralized ETC subsystems. Controller updates decreased by up to 60\%, indicating a significant reduction in computational load. However, prior knowledge of the interconnected system is necessary in the proposed method, which limits the applicability of this method to a wide variety of engineering industries. ML-based decentralized ETC is also studied in \cite{huo2022adaptive} for nonlinear large-scale decentralized control problems with matched interconnections. A single critic network is used to solve the optimal control problem of HJB for nominal isolated subsystems, which decreases the computational cost and avoids the approximation error caused by the actor network. The critic network is updated via modified gradient descent with an additional stability term, and there is no requirement for the initial stabilizing control.

Additionally, ML-based ETC is applied to systems that are subject to denial of service attacks. For example, in \cite{zhang2021adaptive} an iterative single critic learning framework is used in conjunction with ETC to consider the denial-of-service attack for autonomous driving systems, which effectively balances the frequency and changes in adjusting the vehicle’s control during the running process. A single critic network is designed to approximate the optimal cost function and obtain an HJB solution.

\subsection{Reinforcement Learning-Actor Critic}

Online IRL is applied to nonlinear CT systems with external disturbances via an event-triggered mechanism based on robust constrained control problems in \cite{yang2019event}. The event triggered $ H_{\infty} $ tracking control problem is formulated in \cite{yang2019event} as a two-player zero-sum game with a non-quadratic function for constrained inputs. The $H_{\infty} $ controller provides a robust optimal design for nonlinear systems. An $H_{\infty} $ optimal control problem could be formulated in the zero-sum game, based on Basar and Bernhard’s theory\cite{bacsar2008h}. Solving zero-sum games, which is a min-max optimization problem, is normally more preferable than directly solving the $H_{\infty} $ problem. The solution to the event-triggered condition is approximated through an actor-critic structure and a HJI equation. Event-triggered optimal constraint control is obtained through actor NN, and the optimal cost is evaluated based on ADP through a critic NN. Lyapunov stability is also used to validate the closed-loop system's stability.

In \cite{vamvoudakis2018model}, an infinite-horizon optimal adaptive learning problem is formulated to design the control and triggering mechanism of a model-free system. Based on Q-learning, a model-free approach has been derived which will also guarantee the exclusion of Zeno behaviour. An actor-critic structure is selected to adaptively tune the optimal ETC and Q-function for a model-free system using RL to optimize the problem online in order to minimize cost. The system is validated using Lyapunov stability analysis. 

Similar to \cite{vamvoudakis2018model}, infinite horizon integral control is used in \cite{yang2018adaptive}. The authors begin by converting the event-triggered robust nonlinear control problem into an event-triggered nonlinear optimal control problem by constructing an infinite horizon integral cost for the nominal system, whose dynamics are unknown. Then the robust ETC of the original system can be derived via solving the event-triggered nonlinear optimal control problem. A recurrent NN is used to develop the unknown system dynamics and, using these dynamics, a critic network is proposed using adaptive critic design to find the solution to the ET HJB equation. The event-triggered threshold is considered constant and static. The system is validated using Lyapunov stability to show that the system is UUB to origin for all states.

Nonlinear multiagent systems are studied in \cite{bai2021event} via distributed recursive RL ETC. The RBF NN critic and actor are applied to estimate the long-term strategic utility function and the uncertain dynamics in multiagent systems, respectively. The
multi-gradient recursive strategy is tailored to learn the NN
weights, which avoids the local optimal problem in gradient descent-based methods and decreases the dependence of the initial value. Semi-global UUB of all signals in a multiagent system is proven. Combining RL and ETC improved the energy conservation of multiagent systems by reducing the amplitude of the controller signal and the controller update frequency, respectively.

\begin{figure} [h]
    \centering
   \includegraphics[width=\linewidth]{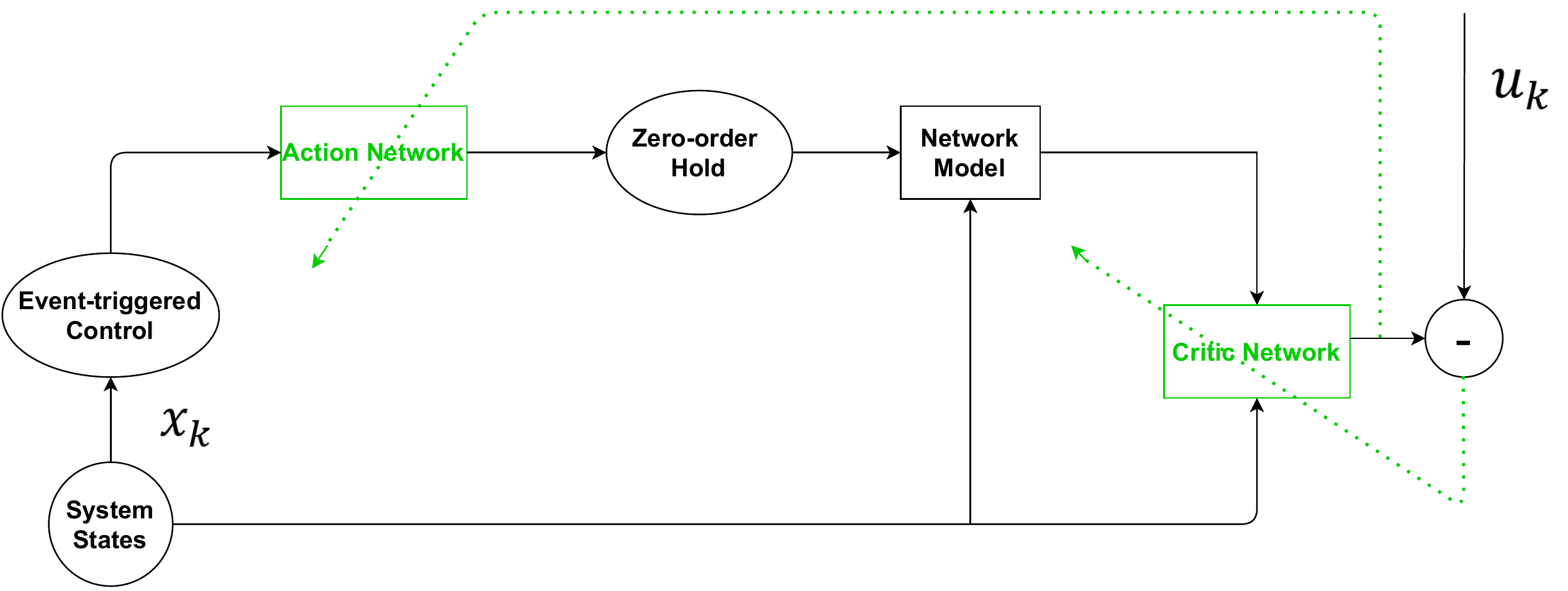}
    \caption{Event-triggered Adaptive Critic Architecture. \cite{wang2019self}}
    \label{fig:MFL}
\end{figure}

A DHDP strategy combined with self-learning optimal regulation for an event-driven adaptive control algorithm has been proposed in \cite{wang2019self}. The DHDP strategy is used to formulate an event-based optimal regulation for discrete time nonlinear systems to reduce the cost. The input-to-state stability (ISS) analysis is proposed for a nonlinear plant. DHDP is a sub-domain of ADP and is used to solve the HJB equations. As shown in Fig. \ref{fig:MFL}, the solid lines denote the flow of state information and the dashed lines denote the back propagation path for both actor and critic networks, where $u_k$ denotes the control input and $x_k$ denotes the state of the system. When compared to the traditional DHDP method, the proposed DHDP technique significantly reduces computation cost and resource utilization while maintaining performance. The ETC has a dynamic threshold.
However, in \cite{wang2019self}, an additional assumption is required that the state norm is bounded by the supremum of the control input norm. For reducing these kinds of assumptions, ETC explainable GDHP is presented for nonlinear discrete-time systems to deal with asymmetric input constraints by integrating an integral function and the actor network in \cite{sun2022event}. An explainable GDHP algorithm is presented to solve the HJB equation online, and the calculations for the derivative of the cost function are relaxed without matrix dimensionality transformations. However, the triggering condition is based on the state feedback scheme, and full-state feedback is required.

In \cite{lu2022event}, using parallel control, a novel event-triggered near-optimal control problem for unknown discrete-time nonlinear systems is studied. To achieve parallel control, the control input is introduced into the feedback system via an augmented nonlinear system with an augmented performance index. The control stability of an augmented nonlinear system is analyzed, and by selecting an appropriate augmented performance index, the optimal control of the augmented system can be viewed as close to optimal control of the original system with the original performance index. Then, a novel ETC based on parallel control and critic-actor network structure is applied without reconstructing unknown systems, thereby avoiding identification errors caused by other learning model approaches. In this method, the initial control input can be set arbitrarily, but control constraints are not considered.

\begin{table*}[h]
\caption{Existing ETC systems proposed in the literature - ML for optimal Control}  \label{tab:comp2}
\scriptsize
\centering          % used for centering table
\begin{tabular}{| c | P{1.2cm} | P{1.2cm} | P{1.4cm} |P{.8cm} |P{.7cm} |P{1.4cm} |P{1.5cm} |P{1.4cm} |P{.8cm} |} 
%\begin{tabular}{| c | c | c |c |c |c |c |c |c|c|}    % centered columns (4 columns)
\hline
 %inserts double horizontal lines
Reference & System Type & Learning Approach & Communication policy & Control policy  & Multi-agent & \seqsplit{Centralized~/~Decentralized~/~Distributed}  & Control Mechanism & \seqsplit{Experimental~Validation}  \\[0.5ex]  % inserts table
%heading
\hline
\cite{cui2019event}& Nonlinear (CT) & RL & \xmark & \cmark & \xmark & \ Cent. & $H_{\infty}$ Control & \xmark\\
\hline
\cite{zhang2016event}& Nonlinear (CT) & CL & \xmark & \cmark & \xmark & \ Cent.& $H_{\infty}$ Control & \xmark\\ 
\hline
\cite{wang2017improving} %\cite{gao2020event}
& Affine Nonlinear (CT) & ACL & \xmark & \cmark & \xmark & \ Cent. & $H_{\infty}$ Control & \xmark\\
\hline
\cite{yang2018event} & Nonlinear (CT) & ACL & \xmark & \cmark & \xmark & \ Cent. & HJB & \xmark\\
\hline
\cite{yang2019event}& Nonlinear (CT) & IRL & \xmark & \cmark & \xmark & \ Cent.  & $H_{\infty}$ Control & \xmark\\
\hline
\cite{zhang2017event}& Nonlinear (CT) & IRL & \xmark & \cmark & \xmark & \ Cent. & HJB & \xmark\\
\hline
\cite{yang2019adaptive} & Nonlinear (CT) & RL \& ER & \xmark & \cmark & \cmark & \ Decent. & HJB & \xmark\\
\hline
\cite{huo2022adaptive} & Nonlinear (CT) & RL & \xmark & \cmark & \cmark & Decent. & HJB & \xmark\\
\hline
\cite{wang2017event} & Affine Nonlinear (CT) & NNs & \xmark & \cmark & \xmark & \ Cent.& HJB & \xmark\\
\hline 
\cite{george2019distributed}& Nonlinear& DL & \cmark & \xmark & \cmark & \ Cent.& --  & \xmark\\
\hline
\cite{yang2020event}& Nonlinear (CT) & ACL & \xmark & \cmark & \xmark & \ Cent.& $H_{\infty} $ control & \xmark\\
\hline 
\cite{qin2022event}& Nonlinear (CT) & RL (Critic-only) & \xmark & \cmark & \xmark &  Cent.& $H_{\infty} $ control & \xmark\\
\hline 
\cite{xue2022event} & Nonlinear (CT) & IRL & \xmark & \cmark & \xmark &  Cent.& Optimal control & \xmark\\
\hline 
\cite{zhao2022goal}& Nonlinear-Discrete & RL & \xmark & \cmark & \xmark & Cent. & State Feedback Optimal control & \xmark\\
\hline 
\cite{vamvoudakis2018model}& Linear & RL  & \cmark & \cmark & \xmark & Cent. & LQR & \xmark\\
\hline 
\cite{yang2018adaptive}& Nonlinear & ACD  & \xmark & \cmark & \xmark & Cent. & Optimal Control & \xmark \\
\hline
\cite{bai2021event} & Nonlinear & ACL  & \xmark & \cmark & \cmark & Distributed & Optimal Control & \xmark \\
\hline
\cite{wang2019self}& Nonlinear & DHDP  & \xmark & \cmark & \xmark & Cent. & Optimal Control & \cmark \\
\hline
\cite{sun2022event}& Nonlinear &  Explainable GHDP  & \xmark & \cmark & \xmark & Cent. & HJB & \xmark\\
\hline
cite{lu2022event}& Nonlinear & ACL  & \xmark & \cmark & \xmark & Cent. & HJB & \xmark\\
\hline
\cite{su2020integral} & Affine Nonlinear & IRL & \xmark & \cmark & \xmark & Cent. & Adaptive Control & \xmark \\
\hline
\cite{zhang2021adaptive} & Nonlinear & RL  & \xmark & \cmark & \xmark & Cent. & HJB & \xmark\\
\hline
\end{tabular}
\end{table*}

\section{Joint Learning and Optimization}\label{sec:ML_JT}
In this section, we review articles, as shown in Table. \ref{tab:comp3}, that aim to achieve two main goals: learning system dynamics, which are unknown, and solving optimization problems for event-triggered optimal control problems. To achieve these goals, an identifier-critic architecture is used by combining RL and NN. In the first step, the identifier NN is applied to learn the system dynamics, and in the second step, the critic NN is utilized to obtain the event-triggered optimal controller \cite{yang2017event}. In some cases, the actor NN is also used. %The Actor-critic architecture is not used in this section, but only the critic approach. 

\subsection{Reinforcement Learning-Critic only}
The model-free RL approach is utilized to simultaneously learn an optimal ETC and the model of the system through an identifier-critic architecture in \cite{yang2017event}. More precisely, the feed-forward NN identifier is used to learn the unknown system dynamics, and the critic NN is used to obtain the event-triggered optimal controller. Standard back-propagation algorithms and e-modification methods \cite{na2019finite} are used together to update the identifier NN. A modified gradient descent method is also used to tune the critic NN. Closed-loop system stability is analysed based on the Lyapunov method, and a single-link robotic arm system is chosen as a nonlinear example for simulation. However, this method is inapplicable to nonlinear systems with non-affine inputs. Similar to \cite{yang2017event}, an event-driven DRL optimization algorithm is developed in \cite{ran2022optimizing} to reduce the energy consumption of data centers. The advantage of ETC over fixed periodic control is the ability to make decisions based on specific events (such as overheating). Event-driven optimization significantly reduces the number of regulatory decisions while assuring adequate system performance. Combining DRL with the high-dimensionality and high dynamics of data centers enables the nonlinear, dynamic aspects of the IT workload and thermal process to be captured. It is demonstrated that event-driven DRL can detect events more effectively, reduce regulatory decisions by 70\% to 95\%, and achieve comparable or even greater energy efficiency. The results of \cite{ran2022optimizing} are compared to \cite{ran2019deepee} and \cite{sun2015event}.

$ H_{\infty} $ event-driven control design based on ACL has been developed to deal with the data-based optimization for a class of unknown nonlinear systems in \cite{wang2017mixed}. A two-player zero-sum differential game adaptive critic controller is designed by combining the event-driven design formulation with a data driven learning identifier used to formulate a nonlinear $ H_{\infty} $ control problem. The unknown dynamics of the plant are learned using a NN-based data-driven design. A unique critic network is considered to solve the event-driven HJI equation. However, disturbance updating is in the time-driven mechanism, which will necessitate choosing the prescribed level of disturbance attenuation appropriately to keep the event-triggering threshold non-negative state dependent and monotonically increasing. The system is validated using Lyapunov stability analysis.

In \cite{zhang2021event}, a neuro-dynamic programming-based ETC method for unknown non-affine nonlinear systems with input constraints is presented. Similar to \cite{yang2017event,wang2017mixed}, a NN identifier is created to discover the unknown system dynamics given input constraints. The value function for solving the event-triggered HJB equation is then approximated using a critic NN. The ETC method can decrease computational load, communication expenses, and bandwidth. This method is applicable to both affine and non-affine systems employing NN identifiers with measurable input and output data.

In \cite{huo2021adaptive}, a decentralized ETC problem is studied for a class of constrained nonlinear interconnected systems. By assigning a distinct cost function to each restricted auxiliary subsystem, the control problem is transformed into the selection of optimal control policies. An event-triggered HJB solution has rendered the system stable and UUB. Utilizing an identifier-critic network architecture relaxes the system's dynamic constraints. An identifier network and a critic network are utilized to identify unknown internal dynamics and approximate optimal cost functions, respectively. Optimizing the weights of the critic network using gradient descent. Combining ETC and RL results in less data transfer and enhanced system performance (less control cost and shorter convergence time).

Some data-driven model research \cite{xu2021single,liu2022data} has been conducted based on constructing models with recurrent neural networks (RNNs) for completely unknown nonlinear systems in order to eliminate identification error and respond quickly to dynamic system changes in system identification. In \cite{xu2021single}, a data-driven model based on RNNs is developed to construct the system uncertainties, including the drift dynamics and the input gain matrix. In the data-driven model, the modeling error caused by NN approximation is eliminated by including a compensation term. A critic NN can approximate the solution of the HJB equation, which significantly simplifies the ACL implementation architecture. In their problem, the authors of \cite{liu2022data} incorporated input constraints and external disturbances to extend the work in \cite{xu2021single}.
By developing an integral Bellman equation in IRL, the authors of \cite{xue2021event} eliminate the system identification procedure. The proposed IRL makes the algorithm suitable for systems whose drift dynamics are unknown. The ETC ADP technique for tracking control of partially unknown systems with constraints and uncertainties is developed. After constructing an augmented function, the optimal tracking control problem with uncertainty is transformed into the optimal regulation of the nominal augmented system with a discounted value function; consequently, the requirement for partial system knowledge is relaxed through the use of IRL. The critic and actor NNs are used, the learning of NN weights is event-triggered, and the initial admissible control requirement is relaxed. However, this method cannot be applied to systems with unmatched uncertainty

\subsection{Reinforcement Learning - Actor Critic}

 An event-triggered HDP $\lambda$ optimal control strategy for nonlinear discrete time systems with unknown dynamics has been developed in \cite{li2021event}. Iteratively, HDP $\lambda$ takes into account a parameter for long-term prediction, the $\lambda$. Although long-term prediction increases accuracy and accelerates the rate of learning, it poses a formidable challenge to control systems with limited bandwidth and computational units. Therefore, ETC ensures system stability and reduces the need for computation and communication. ACI structure or model-actor–critic NN structure is utilized, in which the model NN or identifier NN evaluates the system state in order to obtain $\lambda$-return of the current time target value. Then, actor and critic NN are employed to approximate the event-triggered optimal control signal and the one-step return value, respectively. The Lyapunov approach is used to ensure the UUB stability of the system and the absence of NN weight errors \cite{li2021event}.

In \cite{tan2019event}, an event triggered distributed $ H_{\infty} $ constrained control problem for physically interconnected large-scale partially unknown systems with constrained-input and external disturbance is studied. Using an event-triggered feed-forward control policy, the control of physically interconnected large-scale systems is transformed into equivalent event-triggered control of decoupled multiagent systems. This method has the advantage of learning the solution to the HJI equation by combining the NNs of the critic, identifier, actor, and disturber into one. By omitting three NNs for each agent in a multiagent system, computational complexity and resources are significantly decreased.
\vspace{0.3cm}

\begin{table*}[h]
\caption{Existing ETC systems proposed in the literature - ML for Joint Learning and Optimization}  \label{tab:comp3}
\scriptsize
\centering          % used for centering table
\begin{tabular}{| c | P{1.2cm} | P{1.2cm} | P{1.4cm} |P{.8cm} |P{.7cm} |P{1.4cm} |P{1.5cm} |P{1.4cm} |P{.8cm} |} 
%\begin{tabular}{| c | c | c |c |c |c |c |c |c|c|}    % centered columns (4 columns)
\hline
 %inserts double horizontal lines
Reference & System Type & Learning Approach & Communication policy & Control policy  & Multi-agent & \seqsplit{Centralized~/~Decentralized~/~Distributed}  & Control Mechanism & \seqsplit{Experimental~Validation}  \\[0.5ex]  % inserts table
%heading
\hline
\cite{yang2017event} & Affine Nonlinear& RL & \xmark & \cmark & \xmark & Cent. & HJB &\xmark \\
\hline
\cite{wang2017mixed}& Nonlinear (CT) & RL & \xmark & \cmark & \xmark & Cent. & $H_{\infty}$ Control & \xmark \\
\hline
\cite{zhang2021event}& Non-affine nonlinear & RL & \xmark & \cmark & \xmark & Cent. & HJB & \xmark \\
\hline
\cite{li2021event}& Nonlinear & HDP  & \xmark & \cmark & \xmark & Cent. & Optimal Control & \xmark \\
\hline
\cite{tan2019event}& Nonlinear & RL  & \xmark & \cmark & \cmark & Distributed & $H_{\infty}$ Control & \xmark \\
\hline
\cite{ran2022optimizing} & Nonlinear & DRL  & \xmark & \cmark & \xmark & Cent. & Optimal Control & \xmark \\
\hline
\cite{huo2021adaptive} & Nonlinear Interconnected & RL  & \xmark & \cmark & \xmark & Decent. & HJB & \xmark \\
\hline
\cite{xu2021single} & Nonlinear (CT) & RL \& RNN  & \xmark & \cmark & \xmark & Cent. & HJB & \xmark \\
\hline
\cite{liu2022data} & Nonlinear (CT) & RL \& RNN   & \xmark & \cmark & \xmark & Cent. & HJB & \xmark \\
\hline
\cite{xue2021event} & Affine Nonlinear (CT) & IRL \& NN  & \xmark & \cmark & \xmark & Cent. & Optimal Control & \xmark \\
\hline
\end{tabular}
\end{table*}

\section{Discussion and Open Issues}\label{sec:future_works}
Based on our review of the literature, we can identify a number of open issues and challenges for ML-based ETC/STC systems. We outline some key issues in the following and suggest approaches to address them.

\subsection{Communication Errors}
Reliable real-time data transmission is critical for wireless automation, as it requires real-time system state information from remote observers to determine appropriate control actions. Network-induced packet errors and loss, as well as long and variable communication delays, often occur in wireless communication networks. This is caused by erroneous wireless channels, contention in multi-access wireless communication~\cite{Rap_2001} and packet re-transmissions to reduce packet error rates. In particular, quantization errors, communication delays, and packet loss can cause instability in closed-loop control systems. The existing works on ML-based event-driven control in our comprehensive review in sections~\ref{sec:ML_DL}, ~\ref{sec:ML_OP}, and ~\ref{sec:ML_JT} have not considered the impact of network-induced imperfections in their learning algorithms. In other words, they assume a perfect communication scenario in the sensor-controller communication link and the controller-actuator communication link.

Ignoring network-induced imperfections makes the current results of ML-based event-driven control superficial for real-world applications. Therefore, developing a framework that considers relevant network-induced imperfections is necessary by extending and integrating current results. To cope with communication delays and packet dropouts, several measures need to be taken into account, including: (i) building new data sets; (ii) adapting learning techniques based on the imperfections; and (iii) developing strategies to tackle packet loss and delay. In the following, we present the impact of communication imperfections on ML-based event-driven control in more detail. 

\subsubsection{Packet Loss}
The majority of prior ML-ETC research has assumed that information transmitted by a sender is always successfully received by the receiver. In practice, however, this is not the case. If a packet is lost during transmission from the sensor to the controller or from the controller to the actuator, the ML-ETC will be unaware of the current state. Numerous strategies can be used to mitigate the effect of packet loss in ML-ETC systems. One possibility is to predict the lost information (state) and then use the predicted states to obtain the control input. Another approach to mitigating packet loss is through appropriate error correction design, which can be accomplished via forward error correction (FEC) or backward error correction (BEC) (a.k.a. Automatic Repeat reQuest, ARQ). However, BEC, which is based on re-transmissions, may lead to undesirable communication delays. For example, authors in \cite{cheng2020deeprs} proposed Deep Reed-Solomon (DeepRS) coding, as a novel FEC algorithm which predicts packet loss using deep NNs to determine the amount of redundant packets. While there is research in the literature attempting to combine event-based control and ML in order to deal with packet loss, the majority of these works make assumptions that limit the applicability of the proposed methods. For instance, \cite{lehmann2012event} extend event based state-feedback control to cope with communication delays and packet losses. The maximum tolerable communication delay bound is found, which guarantees the event-based state-feedback control is stable. The results are shown for a communication link with additional packet losses. However, the paper assumes that the dynamics of the plant are considered to be accurately known, the states are measurable, and the communication delay is bounded, which limits the applicability of this method. From a control perspective, robust controllers \cite{xue2019robust,shi2020robust} and MPC \cite{ibrahim2022delay} are well-known to be robust against packet loss and delay in NCSs. However, the performance of these methods usually depends on the dynamic model accuracy and availability, which may strongly vary for real systems. Even if a dynamic model is available, the model of the system might change because of a dynamically changing environment, which can deteriorate the performance of a closed-loop system. This issue can be addressed with the help of ML.

\subsubsection{Network Delays}
As previously stated, existing ML-based ETC methods have not considered communication delays in their problem formulation. There are three types of delays in networked control systems, sensor–controller delays, controller–actuator delays, and controller processing delays. In control theory, these delays cause phase shifts that limit the control bandwidth and affect closed-loop stability \cite{park2017wireless}. In order to overcome the pernicious effects of delay on closed-loop systems, ML can be used in various ways. For example, the average end-to-end delay in communication networks can be modeled accurately using NNs, resulting in improved control with sufficient knowledge of delay uncertainties \cite{mestres2018understanding}. ML can be used for learning models to cope with various uncertainties, such as delays and packet loss. For example, in \cite{jang2019networked}, an MPC is designed for Unmanned Aerial Vehicles (UAVs) and a GP is applied to learn an unknown nonlinear model, whereas \cite{yoo2017learning} also applied a GP-based approach to compensate for random communication delays, which is independent of the UAV's dynamic model. In fact, the pattern of network-induced effects is learned. While the literature presents ML algorithms for learning a system's model and making it robust against delay, delay is not taken into account in ML-ETC methods.

In \cite{wangnetwork}, the authors considered delays only in the filtering phase and not in the control phase. An event-triggered $ {H}_{\infty}$ filter is presented for the description of a Markovian jump system, which considers network-induced delay with the disturbance and an unknown nonlinear perturbation. A NN based on back propagation is used to dynamically adjust the communication threshold to reduce the burden of the network communication. A novel $ {H}_{\infty}$ filtering error system model is used to cope with delays. The NN-based event-triggered scheme is compared with the traditional event-triggered scheme, and the advantage of adjusting the communication threshold dynamically to save more limited communication bandwidth is proven in simulation results. Future research could model delay as stochastic coupled leakage time-varying delays and develop a relaxed Lyapunov–Krasovskii functional for studying the delayed system \cite{cai2021fuzzy}. Similar efforts have been made in \cite{cai2021dissipative} to develop a novel Lyapunov-Krasovskii functional and reveal all intrinsic relationships between time delay and sampling interval in the system.
For example, memory event-triggerd $ {H}_{\infty}$ output feedback control for neural networks with mixed delays including discrete and distributed delay problem is considered in \cite{yan2021memory}. The communication delay among neurons is modeled as a distributed delay term with a kernel representing the probability density and the integral term resulting from the proposed memory event-triggered system can be considered as a second distributed delay term. For designing an event-triggered $ {H}_{\infty}$ controller, the Lyapunov–Krasovskii functional with the distributed delay kernel and a generalized integral inequality helped to form linear matrix inequalities.

\subsection{Quantization Error}
Quantization errors occur in many digital systems during the process of converting signals from analog to digital as a result of the transmission of a plant's state information from a sensor to the controller/learning agent. All the ML-based ETC methods reviewed in this survey consider perfect quantization (i.e., errorless quantization), which limits their application to sensitive control systems. Moreover, quantization plays a significant role in event-driven control systems. As mentioned in ETC, an event is triggered by comparing the norm of state or the norm of the state error, which is a function of the plant's real state information. Both of these comparisons are considered based on non-quantized measurements that are assumed to be known with certainty in the papers we reviewed. This assumption might lead to system instability in practical scenarios \cite{garcia2012model}. Therefore, triggering conditions should be devised based on the available quantized state values. For example, a quantization level based ETC algorithm is presented under measurement uncertainties in \cite{zhou2022quantization}. Note that various types of quantizers are available in the literature, such as static, logarithmic, or dynamic quantizers \cite{mastani2021dynamic}. The impact of these methods should be taken into account in future research studying ML-based event driven control.

\subsection{Mobility Aware Communication and Control}
Mobility has a significant impact on real-time and sensitive control applications such as autonomous cars, robots, unmanned aerial vehicles, and vehicle platoons, where the objects are usually highly dynamic. For these applications, ETC can be a potential tool for ensuring real-time and reliable control actions while conserving wireless communication resources. 
 
Current research on ETC either considers static agents or uses a predefined mobility model in which agents can only move in specific directions, complicating the system model's implementation. Because ML can be used to learn a system's behavior through experience, it can be combined with ETC while taking into account the impact of mobility on the system model. Consideration of the mobility model's availability limits the scope of application. To address this issue, a learning technique can be used, and a preliminary attempt has been made in \cite{kolios2015tracking}, where a data-driven mobility model has been developed. Event detection analysis is conducted based on GPS location readings. This mobility model may be expanded further by allowing both the local and mobile hosts to learn about one another's position using ML algorithms and develop the mobility model based on their experience.

%\subsection{Cyber-Physical Systems and the Internet of Things}

%New part
%Cyber-Physical Systems (CPS) are closely related to Internet of Things (IoT). One might even consider IoT as a subset of networked CPS. IoT forms a key foundation for delivering machine-to-machine (M2M) and machine-to-person communication applications at scale \cite{dartmann2019big}. Due to the close relationship between CPS and IoT, we believe it worth to briefly consider how ML, cloud computing, and IoT enables machines to interact with humans, learning and adapting to their wants and needs \cite{carruthers2016internet}. Several directions for ML research in the IoT can be identified, including ML for intelligent sensing and decision making, ML for latency-guaranteed and ultra-reliable communications, ML for cloud and edge computing, and distributed ML. Further information about ML in IoT and CPS can be found in \cite{xu2018survey}. On the other hand, the performance of ETC was validated in a real-world application in the IoT for the manufacture of polycarboxylate superplasticizer. In \cite{yan2018mixed}, a mixed time-triggered and event-triggered industrial controller is designed for industrial real-time control in the IoT for the purpose of reducing complexity and ensuring reliability. Nevertheless, many interesting topics in IoT remain to be explored through the use of ETC, including real-time reliability and reducing communication in industrial wireless communication setups.

\subsection{Scalability} 
%Previous part
%Scalability is one of the important parameter that needs to be considered for IoT and CPS \cite{7123563,5995279}. As the number of agents attached to an IoT system increases, it is challenging to built event triggered control for such a system. For example, in a centralized structure if the number of agents are large and one of the agent becomes unstable, then it may be difficult to execute ETC since it needs to encounter the worst case transient for every agent. Therefore, scalability of the system should be given significant attention so that ETC can be implemented. 
%New part
%IoT entails an intelligent network infrastructure including, that supports a large number of devices. 
Many of the learning algorithms in the works we reviewed here require significant computational resources, even when using event/self-triggered approaches. This will affect scalability in real-world applications, which is impacted by computational efficiency and reduced communication. The large amounts of data generated by many NCS implementations can be beneficial for learning in order to improve the quality of control via ML and large-scale optimization advances. This can be beneficial, in particular in the very networks that link control systems together. In fact, learn-and-adapt network management schemes result in decreased service delays, increased system resilience, and adaptability. However, in general, learning using large amounts of parameters and data can suffer from the curse of dimensionality, negatively affecting scalability. For further information on these issues, readers are referred to \cite{chen2019learning}. 

A few recent articles have attempted to address scalability issues in specific areas such as semi-definite programming by incorporating methods from ML, control, and robotics \cite{majumdar2020recent}. Saving communication bandwidth is also critical in large-scale projects, and as a result, more scalable ETC techniques should be developed. For example, \cite{cheng2018fully} developed the distributed event-triggered consensus problem for linear multiagent networks. The proposed adaptive event-based protocol is fully distributed and scalable, as it is not reliant on any global information about the network graph or its scale\cite{cheng2018fully}. Event-triggered consensus of linear multiagent systems on undirected graphs is developed with no need to know the precise Laplacian of the communication graph, which keeps the protocol scalable and distributed \cite{li2020consensus}. Scalability should be taken into account in ML-based ETC literature, and additional research is necessary.

\subsection{Cloud/Edge Computing} 
In ML-based ETC systems, agents will often need to execute sophisticated ML and control algorithms. In particular, ML algorithms can be computationally extensive due to their complex nature. Moreover, for an ML-based ETC system, these computations need to be performed in real-time. Due to hardware constraints, it may not always be feasible to perform the ML tasks on an individual agent's hardware platform. One feasible solution is to offload parts of or all of the computational tasks of an ML-based ETC system to cloud or edge nodes \cite{7488250,8016573}. Those nodes can perform the computation and transmit the necessary information back to the agents to assist in the decision-making process. This also relates to the issue of scalability pointed out above. Research needs to be carried out on the aspects of using cloud/edge computing to assist computation in ML-based ETC systems. For example, issues such as the assignment of different tasks to different computing components and locations or communication policy between the agents and cloud/edge nodes need to be thoroughly investigated.

\subsection{Joint Learning of System and Network Models}
\begin{figure} [h]
    \centering
   \includegraphics[width=\linewidth]{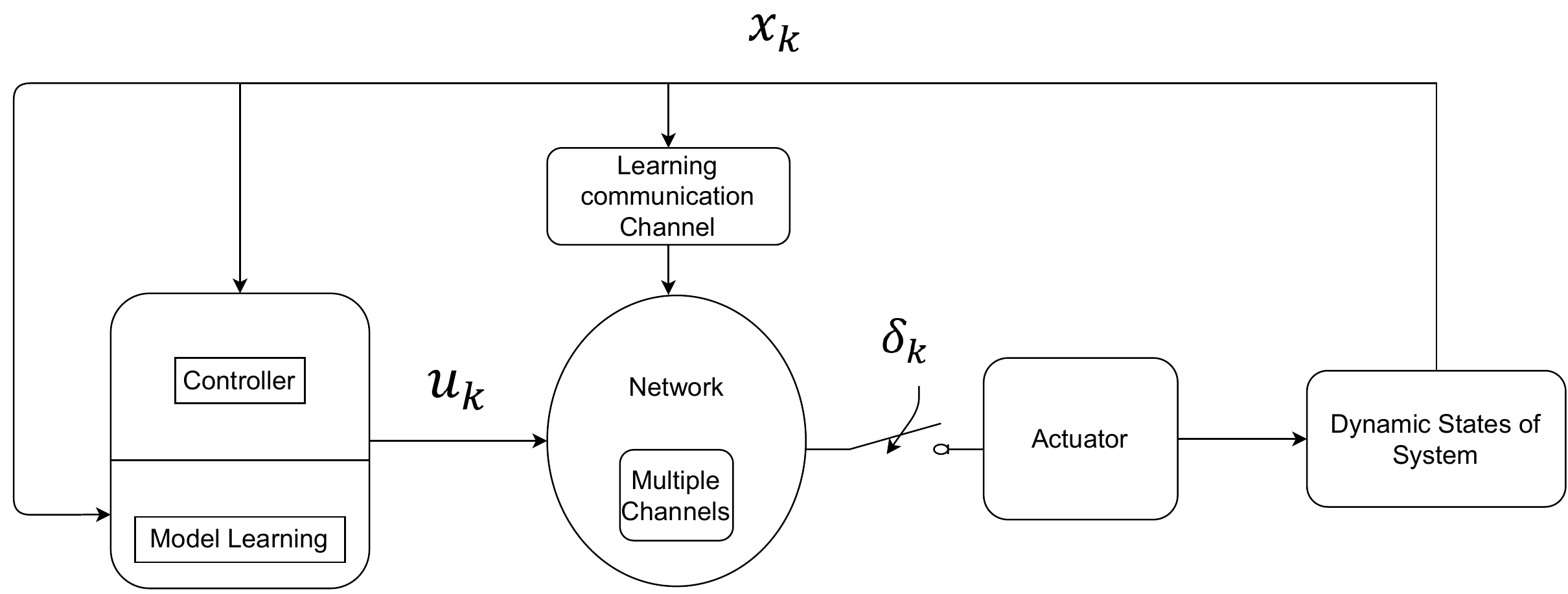}
    \caption{Joint Learning of System and Network Model \cite{redder2019deep}.}
    \label{fig:joint learning}
\end{figure}

Managing the wireless network can play a pivotal role when the control actions are performed over wireless channels. In a real-world scenario, both the system model and the network model may change rapidly. The significance of learning the system model has been understood, and several attempts to use different ML techniques to learn the dynamics of the system model (as discussed earlier) have been presented. Still, the literature assumes that the network model is perfect and always available to the controller, which may not be the case in a practical scenario. The network model also needs to be learned in real time to achieve the best performance for ETC. A preliminary attempt has been made in \cite{redder2019deep}, in which DRL is used to learn the communication network dynamics rather than the plant model, as shown in Fig. \ref{fig:joint learning}, where $x_k$ represents the state of the communication channel, $u_k$ represents the input from the learning agent to the channel, and $\delta_k$ represents the event-triggered threshold.

In large scale NCSs, the number of subsystems may be distributed over a wide area \cite{zhang2017distributed}. The effect of controller awareness on large-scale NCS scheduling decisions is discussed separately in \cite{redder2019deep}. This is the first publication on transmission scheduling for control signals over shared communication channels. The authors use DRL-based iterative resource allocation (DIRA), in which the DIRA uses system state information and performance feedback (control cost evaluations) to achieve optimal control and optimized resource allocation. Further, DIRA can adapt to a given control policy that allows for such performance feedback. The proposed framework does not require a network model, and it implicitly learns the network parameters using DRL. This work can be further extended by using DIRA for state estimation and scheduling of the sensor-controller link along with a time-varying controller-actuator link.

\subsection{Energy Efficient ML-based ETC}

ETC considers a threshold to trigger control actions, resulting in an aperiodic system that is capable of saving computation and communication resources \cite{ge2021dynamic}. ETC can occasionally achieve higher performance with a lower sampling frequency than time-driven control  \cite{li2021event}. By combining ETC and ML, it becomes more robust to disturbances and uncertainties and can potentially be more energy efficient.

In \cite{yoo2019event}, event-triggered MPC is combined with SL to make it more adaptive to uncertainties and robust to state estimation errors. While standard MPC and event-triggered MPC do not account for uncertainties, which can result in tracking errors, learning-based event-triggered MPC can achieve accurate tracking results comparable to standard MPC. The simulation results show that triggering instances can be reduced when using learning-based event-triggered MPC versus event-triggered MPC, highlighting the critical role that ML can play for energy savings. In \cite{baurnann2019event}, model learning is used in an ETL framework when the existing model is not accurate. The benefits of learning system dynamics are demonstrated through a numerical study. After learning, better tracking performance and control signals are observed in comparison to the state prior to learning. Additionally, learning leads to an increase in intercommunication time, resulting in decreased communication. 
Therefore, ML-based ETC can be more energy efficient while maintaining accuracy. While the literature suggests that ETC can reduce communication and lead to energy savings, we believe that applying ML to ETC can result in even greater control accuracy and communication reductions.

\subsection{Self-triggered Control}

As ETC is reactive in nature, it continuously monitors the system's states and triggers when the system deviates too far from a predefined threshold \cite{heemels2012}. ETC requires extra hardware resources for continuous state monitoring, which increases the cost. This is a significant cost in large-scale settings. Another issue with ETC is that it requires full state information at all times, necessitating that the system is more robust throughout the execution of time control actions \cite{heemels2012}. To overcome this, STC can be a better aperiodic triggering mechanism because it is proactive in nature. It calculates the next triggering time at the current time instance, and in-between these instances it remains idle \cite{tiberi2012simple, heemels2012, tiberi2013simple}. It does not require full state information all the time, but rather at triggering instances. This property of STC makes it more suitable for combining with ML algorithms in order to avoid excessive learning of the system model. As discussed earlier in this article, ML is used to learn the dynamic model of the system or to optimize performance, and its combination with ETC also utilizes scarce resources, as ETC requires state information all the time. Therefore, replacing ETC with STC in combination with ML can be extremely beneficial, as it reduces resource waste. 

A preliminary attempt has been made in \cite{hashimoto2019learning}, where the system dynamics are too difficult to obtain due to the system's complexity, necessitating optimal learning via STC-based learning. STC does not require the learning agent to constantly learn the system's model. This STC learning technique can be extended to learn the dynamics required for optimal system control. STC can also be used to combine various ML techniques mentioned earlier to achieve optimal utilization of resources.

\subsection{Security Issues}
In a multiagent system, or NCS, where different nodes communicate with each other, security can be a critical issue. Specifically, security and privacy will be a fundamental challenge to the adoption of large-scale NCSs. Recently, researchers have combined ML and event-triggered communication/control to study fault detection and fault-tolerant control. ML techniques are used to improve the recognition, detection, diagnosis, and prediction accuracy of fault features \cite{cirrincione2020shallow}. The most effective methods for feature classification are deep NNs, recurrent NNs, and conventional NNs. Additionally,  an event-driven approach is used to trigger fault detection and localization in order to improve transmission efficiency \cite{habbouche2021bearing}. From a fault-tolerant control perspective, the authors of \cite{zhang2020composite} used ETC laws to effectively reduce the network transmission load from the controller to the actuators, and they used neural adaptive laws to compensate for unknown actuator faults online. Similar research is being conducted in \cite{zhang2018learning} to address the attitude control problem for spacecraft against actuator faults. ML can also be used to design better control and communication mechanisms that can prevent data injection attacks. A major concern for the traditional training process is privacy, which the nodes may not want to compromise on by sharing training samples. Federated learning, which emerged in recent years to address the privacy and communication overhead issues associated with the training of ML models, has attracted extensive research interest for enhanced wireless networks \cite{Park19FL, Chen2020FL}. Federated learning may play a vital role in the design of future NCS systems. It is worth noting that using ETC for security issues requires smart ETC algorithms. For example, in \cite{zhao2021secure}, a novel event-triggered scheme has been developed, which is smarter and more flexible with features of avoiding chaotic triggering, increasing triggering exponentially, linear compensation, and linear triggering.

%For instance, in \cite{habbouche2021bearing}, variational mode decomposition is used to detect faults and then trigger diagnosis and classification of faults via a conventional NN in an automated and intelligent manner. When a fault is detected, an event-triggered mode of fault diagnosis is activated.

\section{Conclusions} \label{Sect:Conclusions}
This article provides a survey of current ML techniques combined with ETC. We begin our discussion by highlighting the challenge of scarce bandwidth resources available to NCSs and how event-triggered communication can address this challenge. Furthermore, we reviewed various articles that discuss the limitations of implementing ETC for practical NCSs and potential solutions. The majority of the literature indicates that the availability of the model available to the controller is one of the most significant challenges in implementing ETC for NCSs. By learning the entire model or portions of a model, ML is a key technique to address the problems of changing dynamics in practical NCSs. Based on the application of ML in the ETC literature, we classify articles into three groups: dynamic model learning, ML for optimal performance, and joint model learning and optimization. While the literature discusses a variety of ML techniques, ML-based ETC appears to rely primarily on SL, NN, and (deep) RL approaches. Although ML-based ETC approaches have demonstrated promising results in addressing the various challenges outlined here, there is still scope to enhance existing ML approaches further or develop new solutions to address existing challenges. Among them, we highlighted how ML can be used to address issues such as learning the network as well as the system model or how the movement of agents affects the model. We concluded by proposing possible solutions to several of these open issues.

%{\color{brown}An important, related topic, to ML for ETC in NCSs, which we did not address, is security and privacy in networked control systems, particularly those connected to the Internet. We may address this in a future review article.}      

\bibliographystyle{unsrt}
\bibliography{references}

\begin{thebibliography}{100}

\bibitem{8434710}
C.~{Sun}, G.~{Cembrano}, V.~{Puig}, and J.~{Meseguer}.
\newblock Cyber-physical systems for real-time management in the urban water
  cycle.
\newblock In {\em 2018 International Workshop on Cyber-physical Systems for
  Smart Water Networks (CySWater)}, pages 5--8, 2018.

\bibitem{broo2021cyber}
Didem~G{\"u}rd{\"u}r Broo, Ulf Boman, and Martin T{\"o}rngren.
\newblock Cyber-physical systems research and education in 2030: Scenarios and
  strategies.
\newblock {\em Journal of Industrial Information Integration}, 21:100192, 2021.

\bibitem{khujamatov2021iot}
Halim Khujamatov, Ernazar Reypnazarov, Doston Khasanov, and Nurshod Akhmedov.
\newblock Iot, iiot, and cyber-physical systems integration.
\newblock In {\em Emergence of Cyber Physical System and IoT in Smart
  Automation and Robotics}, pages 31--50. Springer, 2021.

\bibitem{pivoto2021cyber}
Diego~GS Pivoto, Luiz~FF de~Almeida, Rodrigo da~Rosa~Righi, Joel~JPC Rodrigues,
  Alexandre~Baratella Lugli, and Antonio~M Alberti.
\newblock Cyber-physical systems architectures for industrial internet of
  things applications in industry 4.0: A literature review.
\newblock {\em Journal of manufacturing systems}, 58:176--192, 2021.

\bibitem{gupta2009networked}
Rachana~Ashok Gupta and Mo-Yuen Chow.
\newblock Networked control system: Overview and research trends.
\newblock {\em IEEE transactions on industrial electronics}, 57(7):2527--2535,
  2009.

\bibitem{mazo2008event}
Manuel Mazo and Paulo Tabuada.
\newblock On event-triggered and self-triggered control over sensor/actuator
  networks.
\newblock In {\em 2008 47th IEEE Conference on Decision and Control}, pages
  435--440. IEEE, 2008.

\bibitem{guinaldo2016distributed}
Mar{\'\i}a Guinaldo, Jos{\'e} S{\'a}nchez, Raquel Dormido, and Sebasti{\'a}n
  Dormido.
\newblock Distributed control for large-scale systems with adaptive
  event-triggering.
\newblock {\em Journal of the Franklin Institute}, 353(3):735--756, 2016.

\bibitem{tiberi2013simple}
U~Tiberi and Karl~Henrik Johansson.
\newblock A simple self-triggered sampler for perturbed nonlinear systems.
\newblock {\em Nonlinear Analysis: Hybrid Systems}, 10:126--140, 2013.

\bibitem{peng2018survey}
Chen Peng and Fuqiang Li.
\newblock A survey on recent advances in event-triggered communication and
  control.
\newblock {\em Information Sciences}, 457:113--125, 2018.

\bibitem{liu2018co}
Shichao Liu, Wensheng Luo, and Ligang Wu.
\newblock Co-design of distributed model-based control and event-triggering
  scheme for load frequency regulation in smart grids.
\newblock {\em IEEE Transactions on Systems, Man, and Cybernetics: Systems},
  50(9):3311--3319, 2018.

\bibitem{solowjow2018event}
Friedrich Solowjow, Dominik Baumann, Jochen Garcke, and Sebastian Trimpe.
\newblock Event-triggered learning for resource-efficient networked control.
\newblock In {\em 2018 Annual American Control Conference (ACC)}, pages
  6506--6512. IEEE, 2018.

\bibitem{baumann2018deep}
Dominik Baumann, Jia-Jie Zhu, Georg Martius, and Sebastian Trimpe.
\newblock Deep reinforcement learning for event-triggered control.
\newblock In {\em 2018 IEEE Conference on Decision and Control (CDC)}, pages
  943--950. IEEE, 2018.

\bibitem{bhasin2012actor}
S~Bhasin, R~Kamalapurkar, M~Johnson, KG~Vamvoudakis, FL~Lewis, and WE~Dixon.
\newblock An actor--critic--identifier architecture for adaptive approximate
  optimal control.
\newblock In {\em Reinforcement Learning and Approximate Dynamic Programming
  for Feedback Control}, pages 258--280. Wiley, 2012.

\bibitem{solowjow2020event}
Friedrich Solowjow and Sebastian Trimpe.
\newblock Event-triggered learning.
\newblock {\em Automatica}, 117:109009, 2020.

\bibitem{baurnann2019event}
Dominik Baumann, Friedrich Solowjow, Karl~Henrik Johansson, and Sebastian
  Trimpe.
\newblock Event-triggered pulse control with model learning (if necessary).
\newblock In {\em 2019 American Control Conference (ACC)}, pages 792--797.
  IEEE, 2019.

\bibitem{yoo2019event}
Jaehyun Yoo and Karl~H Johansson.
\newblock Event-triggered model predictive control with a statistical learning.
\newblock {\em IEEE Transactions on Systems, Man, and Cybernetics: Systems},
  2019.

\bibitem{sahoo2015neural}
Avimanyu Sahoo, Hao Xu, and Sarangapani Jagannathan.
\newblock Neural network-based event-triggered state feedback control of
  nonlinear continuous-time systems.
\newblock {\em IEEE Transactions on Neural Networks and Learning Systems},
  27(3):497--509, 2015.

\bibitem{schluter2019event}
Henning Schluter, Friedrich Solowjow, and Sebastian Trimpe.
\newblock Event-triggered learning for linear quadratic control.
\newblock {\em arXiv preprint arXiv:1910.07732}, 2019.

\bibitem{li2018adaptive}
Yuan-Xin Li and Guang-Hong Yang.
\newblock Adaptive neural control of pure-feedback nonlinear systems with
  event-triggered communications.
\newblock {\em IEEE transactions on neural networks and learning systems},
  29(12):6242--6251, 2018.

\bibitem{li2022event}
Jing Li, Han Liu, Zhaohui Zhang, Xiaobo Li, and Xiaoli Yang.
\newblock Event-triggered adaptive nn tracking control with dynamic gain for a
  class of unknown nonlinear systems.
\newblock {\em Neurocomputing}, 467:292--299, 2022.

\bibitem{chen2021model}
Zhongyu Chen, Ben Niu, Xudong Zhao, Liang Zhang, and Ning Xu.
\newblock Model-based adaptive event-triggered control of nonlinear
  continuous-time systems.
\newblock {\em Applied Mathematics and Computation}, 408:126330, 2021.

\bibitem{liu2021neural}
Lei Liu, Xiangsheng Li, Yan-Jun Liu, and Shaocheng Tong.
\newblock Neural network based adaptive event trigger control for a class of
  electromagnetic suspension systems.
\newblock {\em Control Engineering Practice}, 106:104675, 2021.

\bibitem{hashimoto2019learning}
Kazumune Hashimoto, Yuichi Yoshimura, and Toshimitsu Ushio.
\newblock Learning self-triggered controllers with gaussian processes.
\newblock {\em arXiv preprint arXiv:1909.00178}, 2019.

\bibitem{gao2020event}
Chuang Gao, Chunlei Zhang, Xiaoping Liu, Huanqing Wang, and Lidong Wang.
\newblock Event-triggering based adaptive neural tracking control for a class
  of pure-feedback systems with finite-time prescribed performance.
\newblock {\em Neurocomputing}, 382:221--232, 2020.

\bibitem{liang2020neural}
Hongjing Liang, Guangliang Liu, Huaguang Zhang, and Tingwen Huang.
\newblock Neural-network-based event-triggered adaptive control of nonaffine
  nonlinear multiagent systems with dynamic uncertainties.
\newblock {\em IEEE Transactions on Neural Networks and Learning Systems},
  32(5):2239--2250, 2020.

\bibitem{wang2022event}
Fenglan Wang and Lijun Long.
\newblock Event-triggered adaptive nn control for mimo switched nonlinear
  systems with non-isps unmodeled dynamics.
\newblock {\em Journal of the Franklin Institute}, 2022.

\bibitem{yu2022event}
Shulan Yu, Jinshu Lu, Guibing Zhu, and Shujie Yang.
\newblock Event-triggered finite-time tracking control of underactuated msvs
  based on neural network disturbance observer.
\newblock {\em Ocean Engineering}, 253:111169, 2022.

\bibitem{guo2019event}
Xinxin Guo, Weisheng Yan, and Rongxin Cui.
\newblock Event-triggered reinforcement learning-based adaptive tracking
  control for completely unknown continuous-time nonlinear systems.
\newblock {\em IEEE Transactions on Cybernetics}, 2019.

\bibitem{george2019distributed}
Jemin George and Prudhvi Gurram.
\newblock Distributed deep learning with event-triggered communication.
\newblock {\em arXiv preprint arXiv:1909.05020}, 2019.

\bibitem{zhang2016event}
Qichao Zhang, Dongbin Zhao, and Yuanheng Zhu.
\newblock Event-triggered $ {H}_{\infty}$ control for continuous-time nonlinear
  system via concurrent learning.
\newblock {\em IEEE Transactions on Systems, Man, and Cybernetics: Systems},
  47(7):1071--1081, 2016.

\bibitem{zhang2017event}
Qichao Zhang and Dongbin Zhao.
\newblock Event-triggered integral reinforcement learning for nonlinear
  continuous-time systems.
\newblock In {\em 2017 IEEE Symposium Series on Computational Intelligence
  (SSCI)}, pages 1--6. IEEE, 2017.

\bibitem{wang2017event}
Ding Wang, Haibo He, Xiangnan Zhong, and Derong Liu.
\newblock Event-driven nonlinear discounted optimal regulation involving a
  power system application.
\newblock {\em IEEE Transactions on Industrial Electronics}, 64(10):8177--8186,
  2017.

\bibitem{cui2019event}
Lili Cui, Wei Qu, Li~Wang, Yanhong Luo, and Zhanshan Wang.
\newblock Event-triggered $ {H}_{\infty}$ tracking control of nonlinear systems
  via reinforcement learning method.
\newblock In {\em 2019 International Joint Conference on Neural Networks
  (IJCNN)}, pages 1--8. IEEE, 2019.

\bibitem{yang2018event}
Xiong Yang and Haibo He.
\newblock Event-triggered robust stabilization of nonlinear input-constrained
  systems using single network adaptive critic designs.
\newblock {\em IEEE Transactions on Systems, Man, and Cybernetics: Systems},
  2018.

\bibitem{yang2020event}
Xiong Yang and Haibo He.
\newblock Event-driven $ {H}_{\infty}$-constrained control using adaptive
  critic learning.
\newblock {\em IEEE Transactions on Cybernetics}, 2020.

\bibitem{qin2022event}
Chunbin Qin, Heyang Zhu, Jinguang Wang, Qiyang Xiao, and Dehua Zhang.
\newblock Event-triggered safe control for the zero-sum game of nonlinear
  safety-critical systems with input saturation.
\newblock {\em IEEE Access}, 10:40324--40337, 2022.

\bibitem{xue2022event}
Shan Xue, Biao Luo, Derong Liu, and Ying Gao.
\newblock Event-triggered integral reinforcement learning for nonzero-sum games
  with asymmetric input saturation.
\newblock {\em Neural Networks}, 2022.

\bibitem{su2020integral}
Hanguang Su, Huaguang Zhang, Shaoxin Sun, and Yuliang Cai.
\newblock Integral reinforcement learning-based online adaptive event-triggered
  control for non-zero-sum games of partially unknown nonlinear systems.
\newblock {\em Neurocomputing}, 377:243--255, 2020.

\bibitem{wang2017improving}
Ding Wang, Haibo He, and Derong Liu.
\newblock Improving the critic learning for event-based nonlinear $
  {H}_{\infty}$ control design.
\newblock {\em IEEE transactions on cybernetics}, 47(10):3417--3428, 2017.

\bibitem{zhao2022goal}
Shangwei Zhao, Jingcheng Wang, Hongyuan Wang, and Haotian Xu.
\newblock Goal representation adaptive critic design for discrete-time
  uncertain systems subjected to input constraints: The event-triggered case.
\newblock {\em Neurocomputing}, 2022.

\bibitem{wang2018learning}
Ding Wang and Derong Liu.
\newblock Learning and guaranteed cost control with event-based adaptive critic
  implementation.
\newblock {\em IEEE transactions on neural networks and learning systems},
  29(12):6004--6014, 2018.

\bibitem{yang2019adaptive}
Xiong Yang and Haibo He.
\newblock Adaptive critic learning and experience replay for decentralized
  event-triggered control of nonlinear interconnected systems.
\newblock {\em IEEE Transactions on Systems, Man, and Cybernetics: Systems},
  2019.

\bibitem{huo2022adaptive}
Yu~Huo, Ding Wang, and Junfei Qiao.
\newblock Adaptive critic optimization to decentralized event-triggered control
  of continuous-time nonlinear interconnected systems.
\newblock {\em Optimal Control Applications and Methods}, 43(1):198--212, 2022.

\bibitem{zhang2021adaptive}
Kun Zhang, Rong Su, Huaguang Zhang, and Yunlin Tian.
\newblock Adaptive resilient event-triggered control design of autonomous
  vehicles with an iterative single critic learning framework.
\newblock {\em IEEE transactions on neural networks and learning systems},
  32(12):5502--5511, 2021.

\bibitem{yang2019event}
Dongsheng Yang, Ting Li, Huaguang Zhang, and Xiangpeng Xie.
\newblock Event-trigger-based robust control for nonlinear constrained-input
  systems using reinforcement learning method.
\newblock {\em Neurocomputing}, 340:158--170, 2019.

\bibitem{vamvoudakis2018model}
Kyriakos~G Vamvoudakis and Henrique Ferraz.
\newblock Model-free event-triggered control algorithm for continuous-time
  linear systems with optimal performance.
\newblock {\em Automatica}, 87:412--420, 2018.

\bibitem{yang2018adaptive}
Xiong Yang and Haibo He.
\newblock Adaptive critic designs for event-triggered robust control of
  nonlinear systems with unknown dynamics.
\newblock {\em IEEE transactions on cybernetics}, 49(6):2255--2267, 2018.

\bibitem{bai2021event}
Weiwei Bai, Tieshan Li, Yue Long, and CL~Philip Chen.
\newblock Event-triggered multigradient recursive reinforcement learning
  tracking control for multiagent systems.
\newblock {\em IEEE Transactions on Neural Networks and Learning Systems},
  2021.

\bibitem{wang2019self}
Ding Wang, Mingming Ha, and Junfei Qiao.
\newblock Self-learning optimal regulation for discrete-time nonlinear systems
  under event-driven formulation.
\newblock {\em IEEE Transactions on Automatic Control}, 2019.

\bibitem{sun2022event}
Bo~Sun and Erik-Jan van Kampen.
\newblock Event-triggered constrained control using explainable global dual
  heuristic programming for nonlinear discrete-time systems.
\newblock {\em Neurocomputing}, 468:452--463, 2022.

\bibitem{lu2022event}
Jingwei Lu, Qinglai Wei, Tianmin Zhou, Ziyang Wang, and Fei-Yue Wang.
\newblock Event-triggered near-optimal control for unknown discrete-time
  nonlinear systems using parallel control.
\newblock {\em IEEE Transactions on Cybernetics}, 2022.

\bibitem{yang2017event}
Xiong Yang, Haibo He, and Derong Liu.
\newblock Event-triggered optimal neuro-controller design with reinforcement
  learning for unknown nonlinear systems.
\newblock {\em IEEE Transactions on Systems, Man, and Cybernetics: Systems},
  2017.

\bibitem{ran2022optimizing}
Yongyi Ran, Xin Zhou, Han Hu, and Yonggang Wen.
\newblock Optimizing data centre energy efficiency via event-driven deep
  reinforcement learning.
\newblock {\em IEEE Transactions on Services Computing}, 2022.

\bibitem{wang2017mixed}
Ding Wang, Chaoxu Mu, Derong Liu, and Hongwen Ma.
\newblock On mixed data and event driven design for adaptive-critic-based
  nonlinear $ {H}_{\infty}$ control.
\newblock {\em IEEE transactions on neural networks and learning systems},
  29(4):993--1005, 2017.

\bibitem{zhang2021event}
Shunchao Zhang, Bo~Zhao, and Yongwei Zhang.
\newblock Event-triggered control for input constrained non-affine nonlinear
  systems based on neuro-dynamic programming.
\newblock {\em Neurocomputing}, 440:175--184, 2021.

\bibitem{huo2021adaptive}
Xin Huo, Hamid~Reza Karimi, Xudong Zhao, Bohui Wang, and Guangdeng Zong.
\newblock Adaptive-critic design for decentralized event-triggered control of
  constrained nonlinear interconnected systems within an identifier-critic
  framework.
\newblock {\em IEEE Transactions on Cybernetics}, 2021.

\bibitem{xu2021single}
Ning Xu, Ben Niu, Huanqing Wang, Xin Huo, and Xudong Zhao.
\newblock Single-network adp for solving optimal event-triggered tracking
  control problem of completely unknown nonlinear systems.
\newblock {\em International Journal of Intelligent Systems}, 36(9):4795--4815,
  2021.

\bibitem{liu2022data}
Shanlin Liu, Ben Niu, Guangdeng Zong, Xudong Zhao, and Ning Xu.
\newblock Data-driven-based event-triggered optimal control of unknown
  nonlinear systems with input constraints.
\newblock {\em Nonlinear Dynamics}, pages 1--19, 2022.

\bibitem{xue2021event}
Shan Xue, Biao Luo, Derong Liu, and Ying Gao.
\newblock Event-triggered adp for tracking control of partially unknown
  constrained uncertain systems.
\newblock {\em IEEE Transactions on Cybernetics}, 2021.

\bibitem{li2021event}
Ting Li, Dongsheng Yang, Xiangpeng Xie, and Huaguang Zhang.
\newblock Event-triggered control of nonlinear discrete-time system with
  unknown dynamics based on hdp ($\lambda$).
\newblock {\em IEEE Transactions on Cybernetics}, 2021.

\bibitem{tan2019event}
Luy~Nguyen Tan.
\newblock Event-triggered distributed h∞ constrained control of physically
  interconnected large-scale partially unknown strict-feedback systems.
\newblock {\em IEEE Transactions on Systems, Man, and Cybernetics: Systems},
  51(4):2444--2456, 2019.

\bibitem{ijaz2018self}
Zohaib Ijaz, Muhammad Tahir, and Sahar Arshad.
\newblock Self-triggered control plane for cognitive radio networks.
\newblock In {\em 2018 IEEE 88th Vehicular Technology Conference (VTC-Fall)},
  pages 1--5. IEEE, 2018.

\bibitem{wang2021neural}
Jianhui Wang, Hongkang Zhang, Kemao Ma, Zhi Liu, and CL~Philip Chen.
\newblock Neural adaptive self-triggered control for uncertain nonlinear
  systems with input hysteresis.
\newblock {\em IEEE Transactions on Neural Networks and Learning Systems},
  2021.

\bibitem{heemels2012}
W.~P. M.~H. {Heemels}, K.~H. {Johansson}, and P.~{Tabuada}.
\newblock An introduction to event-triggered and self-triggered control.
\newblock In {\em 2012 IEEE 51st IEEE Conference on Decision and Control
  (CDC)}, pages 3270--3285, 2012.

\bibitem{ran2020event}
Guangtao Ran, Chuanjiang Li, Hak-Keung Lam, Dongyu Li, and Chunsong Han.
\newblock Event-based dissipative control of interval type-2 fuzzy markov jump
  systems under sensor saturation and actuator nonlinearity.
\newblock {\em IEEE Transactions on Fuzzy Systems}, 2020.

\bibitem{wang2021cooperative}
Shimin Wang, Zhan Shu, and Tongwen Chen.
\newblock Cooperative output regulation with mixed time-and event-triggered
  observers.
\newblock {\em arXiv preprint arXiv:2105.02200}, 2021.

\bibitem{tian2019probabilistic}
Engang Tian, Zidong Wang, Lei Zou, and Dong Yue.
\newblock Probabilistic-constrained filtering for a class of nonlinear systems
  with improved static event-triggered communication.
\newblock {\em International Journal of Robust and Nonlinear Control},
  29(5):1484--1498, 2019.

\bibitem{ge2020dynamic}
Xiaohua Ge, Qing-Long Han, Lei Ding, Yu-Long Wang, and Xian-Ming Zhang.
\newblock Dynamic event-triggered distributed coordination control and its
  applications: A survey of trends and techniques.
\newblock {\em IEEE Transactions on Systems, Man, and Cybernetics: Systems},
  50(9):3112--3125, 2020.

\bibitem{gu2017adaptive}
Zhou Gu, Engang Tian, and Jinliang Liu.
\newblock Adaptive event-triggered control of a class of nonlinear networked
  systems.
\newblock {\em Journal of the Franklin Institute}, 354(9):3854--3871, 2017.

\bibitem{zhao2020hybrid}
Guanglei Zhao, Changchun Hua, and Xinping Guan.
\newblock A hybrid event-triggered approach to consensus of multiagent systems
  with disturbances.
\newblock {\em IEEE Transactions on Control of Network Systems},
  7(3):1259--1271, 2020.

\bibitem{ge2021dynamic}
Xiaohua Ge, Qing-Long Han, Xian-Ming Zhang, and Derui Ding.
\newblock Dynamic event-triggered control and estimation: A survey.
\newblock {\em International Journal of Automation and Computing},
  18(6):857--886, 2021.

\bibitem{chen2020often}
Zhiyong Chen, Qing-Long Han, Yamin Yan, and Zheng-Guang Wu.
\newblock How often should one update control and estimation: review of
  networked triggering techniques.
\newblock {\em Science China Information Sciences}, 63(5):1--18, 2020.

\bibitem{ran2022adaptive}
Guangtao Ran, Chuanjiang Li, Sakthivel Rathinasamy, Chunsong Han, Bohui Wang,
  and Jian Liu.
\newblock Adaptive event-triggered asynchronous control for interval type-2
  fuzzy markov jump systems with cyber-attacks.
\newblock {\em IEEE Transactions on Control of Network Systems}, 2022.

\bibitem{mitchell1997machine}
Thomas~M Mitchell et~al.
\newblock Machine learning, 1997.

\bibitem{koltchinski2000statistical}
V~Koltchinski, Abdallah C.T., M~Ariolay, P.~Dorato, and D.~Panchenko.
\newblock Statistical learning control of uncertain systems: It is better than
  it seems.
\newblock Technical report, University of New Mexico, 2000.

\bibitem{papadimitrakis2022active}
Myron Papadimitrakis and Alex Alexandridis.
\newblock Active vehicle suspension control using road preview model predictive
  control and radial basis function networks.
\newblock {\em Applied Soft Computing}, 120:108646, 2022.

\bibitem{golnaraghi2020predicting}
Sasan Golnaraghi, Osama Moselhi, Sabah Alkass, and Zahra Zangenehmadar.
\newblock Predicting construction labor productivity using lower upper
  decomposition radial base function neural network.
\newblock {\em Engineering Reports}, 2(2):e12107, 2020.

\bibitem{singh2021reinforcement}
Bharat Singh, Rajesh Kumar, and Vinay~Pratap Singh.
\newblock Reinforcement learning in robotic applications: a comprehensive
  survey.
\newblock {\em Artificial Intelligence Review}, pages 1--46, 2021.

\bibitem{peters2005natural}
Jan Peters, Sethu Vijayakumar, and Stefan Schaal.
\newblock Natural actor-critic.
\newblock In {\em European Conference on Machine Learning}, pages 280--291.
  Springer, 2005.

\bibitem{nasir2021review}
Vahid Nasir and Farrokh Sassani.
\newblock A review on deep learning in machining and tool monitoring: methods,
  opportunities, and challenges.
\newblock {\em The International Journal of Advanced Manufacturing Technology},
  115(9):2683--2709, 2021.

\bibitem{nguyen2019machine}
Giang Nguyen, Stefan Dlugolinsky, Martin Bob{\'a}k, Viet Tran, Alvaro
  Lopez~Garcia, Ignacio Heredia, Peter Mal{\'\i}k, and Ladislav Hluch{\`y}.
\newblock Machine learning and deep learning frameworks and libraries for
  large-scale data mining: a survey.
\newblock {\em Artificial Intelligence Review}, 52(1):77--124, 2019.

\bibitem{ashraf2021state}
Nesma~M Ashraf, Reham~R Mostafa, Rasha~H Sakr, and MZ~Rashad.
\newblock A state-of-the-art review of deep reinforcement learning techniques
  for real-time strategy games.
\newblock {\em Applications of Artificial Intelligence in Business, Education
  and Healthcare}, pages 285--307, 2021.

\bibitem{schwenzer2021review}
Max Schwenzer, Muzaffer Ay, Thomas Bergs, and Dirk Abel.
\newblock Review on model predictive control: an engineering perspective.
\newblock {\em The International Journal of Advanced Manufacturing Technology},
  117(5):1327--1349, 2021.

\bibitem{dabney2020temporally}
Will Dabney, Georg Ostrovski, and Andr{\'e} Barreto.
\newblock Temporally-extended $\{$$\backslash$epsilon$\}$-greedy exploration.
\newblock {\em arXiv preprint arXiv:2006.01782}, 2020.

\bibitem{li2021distributed}
Tieshan Li, Weiwei Bai, Qi~Liu, Yue Long, and CL~Philip Chen.
\newblock Distributed fault-tolerant containment control protocols for the
  discrete-time multiagent systems via reinforcement learning method.
\newblock {\em IEEE Transactions on Neural Networks and Learning Systems},
  2021.

\bibitem{mu2019learning}
Chaoxu Mu and Yong Zhang.
\newblock Learning-based robust tracking control of quadrotor with time-varying
  and coupling uncertainties.
\newblock {\em IEEE transactions on neural networks and learning systems},
  31(1):259--273, 2019.

\bibitem{chen2022novel}
Zhe Chen, Wenqian Xue, Ning Li, Bosen Lian, and Frank~L Lewis.
\newblock A novel z-function-based completely model-free reinforcement learning
  method to finite-horizon zero-sum game of nonlinear system.
\newblock {\em Nonlinear Dynamics}, pages 1--20, 2022.

\bibitem{su2021event}
Hanguang Su, Huaguang Zhang, Yanhong Luo, and Qiuye Sun.
\newblock Event-based integral reinforcement learning algorithm for
  non-zero-sum games of partially unknown nonlinear systems.
\newblock In {\em 2021 IEEE 10th Data Driven Control and Learning Systems
  Conference (DDCLS)}, pages 287--292. IEEE, 2021.

\bibitem{wen2021optimized}
Guoxing Wen, Liguang Xu, and Bin Li.
\newblock Optimized backstepping tracking control using reinforcement learning
  for a class of stochastic nonlinear strict-feedback systems.
\newblock {\em IEEE Transactions on Neural Networks and Learning Systems},
  2021.

\bibitem{zhang2021observer}
Shunchao Zhang, Bo~Zhao, Derong Liu, and Yongwei Zhang.
\newblock Observer-based event-triggered control for zero-sum games of input
  constrained multi-player nonlinear systems.
\newblock {\em Neural Networks}, 144:101--112, 2021.

\bibitem{parikh2019integral}
Anup Parikh, Rushikesh Kamalapurkar, and Warren~E Dixon.
\newblock Integral concurrent learning: Adaptive control with parameter
  convergence using finite excitation.
\newblock {\em International Journal of Adaptive Control and Signal
  Processing}, 33(12):1775--1787, 2019.

\bibitem{yasini2015online}
Sholeh Yasini, Ali Karimpour, Mohammad-Bagher Naghibi~Sistani, and Hamidreza
  Modares.
\newblock Online concurrent reinforcement learning algorithm to solve
  two-player zero-sum games for partially unknown nonlinear continuous-time
  systems.
\newblock {\em International Journal of Adaptive Control and Signal
  Processing}, 29(4):473--493, 2015.

\bibitem{vamvoudakis2012online}
Kyriakos~G Vamvoudakis and Frank~L Lewis.
\newblock Online solution of nonlinear two-player zero-sum games using
  synchronous policy iteration.
\newblock {\em International Journal of Robust and Nonlinear Control},
  22(13):1460--1483, 2012.

\bibitem{peng2021online}
Chi Peng and Jianjun Ma.
\newblock Online integral reinforcement learning control for an uncertain
  highly flexible aircraft using state and output feedback.
\newblock {\em Aerospace Science and Technology}, 109:106442, 2021.

\bibitem{granzotto2020finite}
Mathieu Granzotto, Romain Postoyan, Lucian Bu{\c{s}}oniu, Dragan
  Ne{\v{s}}i{\'c}, and Jamal Daafouz.
\newblock Finite-horizon discounted optimal control: stability and performance.
\newblock {\em IEEE Transactions on Automatic Control}, 66(2):550--565, 2020.

\bibitem{bacsar2008h}
Tamer Ba{\c{s}}ar and Pierre Bernhard.
\newblock {\em $ {H}_{\infty}$ optimal control and related minimax design
  problems: a dynamic game approach}.
\newblock Springer Science \& Business Media, 2008.

\bibitem{na2019finite}
Jing Na, Shubo Wang, Yan-Jun Liu, Yingbo Huang, and Xuemei Ren.
\newblock Finite-time convergence adaptive neural network control for nonlinear
  servo systems.
\newblock {\em IEEE Transactions on Cybernetics}, 50(6):2568--2579, 2019.

\bibitem{ran2019deepee}
Yongyi Ran, Han Hu, Xin Zhou, and Yonggang Wen.
\newblock Deepee: Joint optimization of job scheduling and cooling control for
  data center energy efficiency using deep reinforcement learning.
\newblock In {\em 2019 IEEE 39th International Conference on Distributed
  Computing Systems (ICDCS)}, pages 645--655. IEEE, 2019.

\bibitem{sun2015event}
Biao Sun, Peter~B Luh, Qing-Shan Jia, and Bing Yan.
\newblock Event-based optimization within the lagrangian relaxation framework
  for energy savings in hvac systems.
\newblock {\em IEEE Transactions on Automation Science and Engineering},
  12(4):1396--1406, 2015.

\bibitem{Rap_2001}
Theodore Rappaport.
\newblock {\em Wireless Communications: Principles and Practice}.
\newblock Prentice Hall PTR, USA, 2nd edition, 2001.

\bibitem{cheng2020deeprs}
Sheng Cheng, Han Hu, Xinggong Zhang, and Zongming Guo.
\newblock Deeprs: Deep-learning based network-adaptive fec for real-time video
  communications.
\newblock {\em arXiv preprint arXiv:2001.07852}, 2020.

\bibitem{lehmann2012event}
Daniel Lehmann and Jan Lunze.
\newblock Event-based control with communication delays and packet losses.
\newblock {\em International Journal of Control}, 85(5):563--577, 2012.

\bibitem{xue2019robust}
Binqiang Xue, Haisheng Yu, and Mengling Wang.
\newblock Robust $ {H}_{\infty}$ output feedback control of networked control
  systems with discrete distributed delays subject to packet dropout and
  quantization.
\newblock {\em IEEE Access}, 7:30313--30320, 2019.

\bibitem{shi2020robust}
Yuanbo Shi, Jianhui Wang, Xiaoke Fang, Yueyang Huang, and Shusheng Gu.
\newblock Robust mixed ${H}_2/{H}_{\infty}$ control for an uncertain wireless
  sensor network systems with time delay and packet loss.
\newblock {\em International Journal of Control, Automation and Systems}, pages
  1--13, 2020.

\bibitem{ibrahim2022delay}
Amr Ibrahim, Dip Goswami, Hong Li, and Twan Basten.
\newblock Delay-aware multi-layer multi-rate model predictive control for
  vehicle platooning under message-rate congestion control.
\newblock {\em IEEE Access}, 2022.

\bibitem{park2017wireless}
Pangun Park, Sinem~Coleri Ergen, Carlo Fischione, Chenyang Lu, and Karl~Henrik
  Johansson.
\newblock Wireless network design for control systems: A survey.
\newblock {\em IEEE Communications Surveys \& Tutorials}, 20(2):978--1013,
  2017.

\bibitem{mestres2018understanding}
Albert Mestres, Eduard Alarc{\'o}n, Yusheng Ji, and Albert Cabellos-Aparicio.
\newblock Understanding the modeling of computer network delays using neural
  networks.
\newblock In {\em Proceedings of the 2018 Workshop on Big Data Analytics and
  Machine Learning for Data Communication Networks}, pages 46--52, 2018.

\bibitem{jang2019networked}
Dohyun Jang, Jaehyun Yoo, Clark~Youngdong Son, H~Jin Kim, and Karl~H Johansson.
\newblock Networked operation of a uav using gaussian process-based delay
  compensation and model predictive control.
\newblock In {\em 2019 International Conference on Robotics and Automation
  (ICRA)}, pages 9216--9222. IEEE, 2019.

\bibitem{yoo2017learning}
Jaehyun Yoo and Karl~H Johansson.
\newblock Learning communication delay patterns for remotely controlled uav
  networks.
\newblock {\em IFAC-PapersOnLine}, 50(1):13216--13221, 2017.

\bibitem{wangnetwork}
Yuzhong Wang, Tie Zhang, Si~Chen, and Junchao Ren.
\newblock Network-based $ h_{\infty} $ filtering for descriptor markovian jump
  systems with a novel neural network event-triggered scheme.
\newblock {\em Neural Processing Letters}, pages 1--19, 2021.

\bibitem{cai2021fuzzy}
Xiao Cai, Jun Wang, Shouming Zhong, Kaibo Shi, and Yiqian Tang.
\newblock Fuzzy quantized sampled-data control for extended dissipative
  analysis of t--s fuzzy system and its application to wpgss.
\newblock {\em Journal of the Franklin Institute}, 358(2):1350--1375, 2021.

\bibitem{cai2021dissipative}
Xiao Cai, Kaibo Shi, Shouming Zhong, Jun Wang, and Yiqian Tang.
\newblock Dissipative analysis for high speed train systems via
  looped-functional and relaxed condition methods.
\newblock {\em Applied Mathematical Modelling}, 96:570--583, 2021.

\bibitem{yan2021memory}
Shen Yan, Zhou Gu, and Sing~Kiong Nguang.
\newblock Memory-event-triggered h∞ output control of neural networks with
  mixed delays.
\newblock {\em IEEE Transactions on Neural Networks and Learning Systems},
  2021.

\bibitem{garcia2012model}
Eloy Garcia and Panos~J Antsaklis.
\newblock Model-based event-triggered control for systems with quantization and
  time-varying network delays.
\newblock {\em IEEE Transactions on Automatic Control}, 58(2):422--434, 2012.

\bibitem{zhou2022quantization}
Tianwei Zhou, Guanghui Yue, and Ben Niu.
\newblock Quantization level based event-triggered control with measurement
  uncertainties.
\newblock {\em Information Sciences}, 588:442--456, 2022.

\bibitem{mastani2021dynamic}
Elahe Mastani and Mehdi Rahmani.
\newblock Dynamic output feedback control for networked systems subject to
  communication delays, packet dropouts, and quantization.
\newblock {\em Journal of the Franklin Institute}, 358(8):4303--4325, 2021.

\bibitem{kolios2015tracking}
Panayiotis Kolios, Georgios Ellinas, and Christos Panayiotou.
\newblock Tracking trip changes with event triggering.
\newblock In {\em 2015 IEEE 18th International Conference on Intelligent
  Transportation Systems}, pages 529--534. IEEE, 2015.

\bibitem{chen2019learning}
Tianyi Chen, Sergio Barbarossa, Xin Wang, Georgios~B Giannakis, and Zhi-Li
  Zhang.
\newblock Learning and management for internet of things: Accounting for
  adaptivity and scalability.
\newblock {\em Proceedings of the IEEE}, 107(4):778--796, 2019.

\bibitem{majumdar2020recent}
Anirudha Majumdar, Georgina Hall, and Amir~Ali Ahmadi.
\newblock Recent scalability improvements for semidefinite programming with
  applications in machine learning, control, and robotics.
\newblock {\em Annual Review of Control, Robotics, and Autonomous Systems},
  3:331--360, 2020.

\bibitem{cheng2018fully}
Bin Cheng and Zhongkui Li.
\newblock Fully distributed event-triggered protocols for linear multiagent
  networks.
\newblock {\em IEEE Transactions on Automatic Control}, 64(4):1655--1662, 2018.

\bibitem{li2020consensus}
Xianwei Li, Yang Tang, and Hamid~Reza Karimi.
\newblock Consensus of multi-agent systems via fully distributed
  event-triggered control.
\newblock {\em Automatica}, 116:108898, 2020.

\bibitem{7488250}
W.~{Shi}, J.~{Cao}, Q.~{Zhang}, Y.~{Li}, and L.~{Xu}.
\newblock Edge computing: Vision and challenges.
\newblock {\em IEEE Internet of Things Journal}, 3(5):637--646, 2016.

\bibitem{8016573}
Y.~{Mao}, C.~{You}, J.~{Zhang}, K.~{Huang}, and K.~B. {Letaief}.
\newblock A survey on mobile edge computing: The communication perspective.
\newblock {\em IEEE Communications Surveys Tutorials}, 19(4):2322--2358, 2017.

\bibitem{redder2019deep}
Adrian Redder, Arunselvan Ramaswamy, and Daniel~E Quevedo.
\newblock Deep reinforcement learning for scheduling in large-scale networked
  control systems.
\newblock {\em IFAC-PapersOnLine}, 52(20):333--338, 2019.

\bibitem{zhang2017distributed}
Dan Zhang, Sing~Kiong Nguang, and Li~Yu.
\newblock Distributed control of large-scale networked control systems with
  communication constraints and topology switching.
\newblock {\em IEEE Transactions on Systems, Man, and Cybernetics: Systems},
  47(7):1746--1757, 2017.

\bibitem{tiberi2012simple}
Ubaldo Tiberi and Karl~Henrik Johansson.
\newblock A simple self-triggered sampler for nonlinear systems.
\newblock {\em IFAC Proceedings Volumes}, 45(9):76--81, 2012.

\bibitem{cirrincione2020shallow}
Giansalvo Cirrincione, Rahul~Ranjeev Kumar, Ali Mohammadi, Shahin~Hedayati Kia,
  Pietro Barbiero, and Jacopo Ferretti.
\newblock Shallow versus deep neural networks in gear fault diagnosis.
\newblock {\em IEEE Transactions on Energy Conversion}, 35(3):1338--1347, 2020.

\bibitem{habbouche2021bearing}
Houssem Habbouche, Yassine Amirat, Tarak Benkedjouh, and Mohamed Benbouzid.
\newblock Bearing fault event-triggered diagnosis using a variational mode
  decomposition-based machine learning approach.
\newblock {\em IEEE Transactions on Energy Conversion}, 2021.

\bibitem{zhang2020composite}
Guoqing Zhang, Shengjia Chu, Xu~Jin, and Weidong Zhang.
\newblock Composite neural learning fault-tolerant control for underactuated
  vehicles with event-triggered input.
\newblock {\em IEEE Transactions on Cybernetics}, 51(5):2327--2338, 2020.

\bibitem{zhang2018learning}
Chengxi Zhang, Jihe Wang, Dexin Zhang, and Xiaowei Shao.
\newblock Learning observer based and event-triggered control to spacecraft
  against actuator faults.
\newblock {\em Aerospace Science and Technology}, 78:522--530, 2018.

\bibitem{Park19FL}
J.~{Park}, S.~{Samarakoon}, M.~{Bennis}, and M.~{Debbah}.
\newblock Wireless network intelligence at the edge.
\newblock {\em Proceedings of the IEEE}, 107(11):2204--2239, 2019.

\bibitem{Chen2020FL}
M.~Chen, Z.~Yang, W.~Saad, C.~Yin, H.~V. Poor, and S.~Cui.
\newblock A joint learning and communications framework for federated learning
  over wireless networks.
\newblock {\em IEEE Transactions on Wireless Communications, to appear}, 2020.

\bibitem{zhao2021secure}
Can Zhao, Xinzhi Liu, Shouming Zhong, Kaibo Shi, Daixi Liao, and Qishui Zhong.
\newblock Secure consensus of multi-agent systems with redundant signal and
  communication interference via distributed dynamic event-triggered control.
\newblock {\em ISA transactions}, 112:89--98, 2021.

\end{thebibliography}

\begin{IEEEbiography}[{\includegraphics[width=1in,height=1.25in,clip,keepaspectratio]{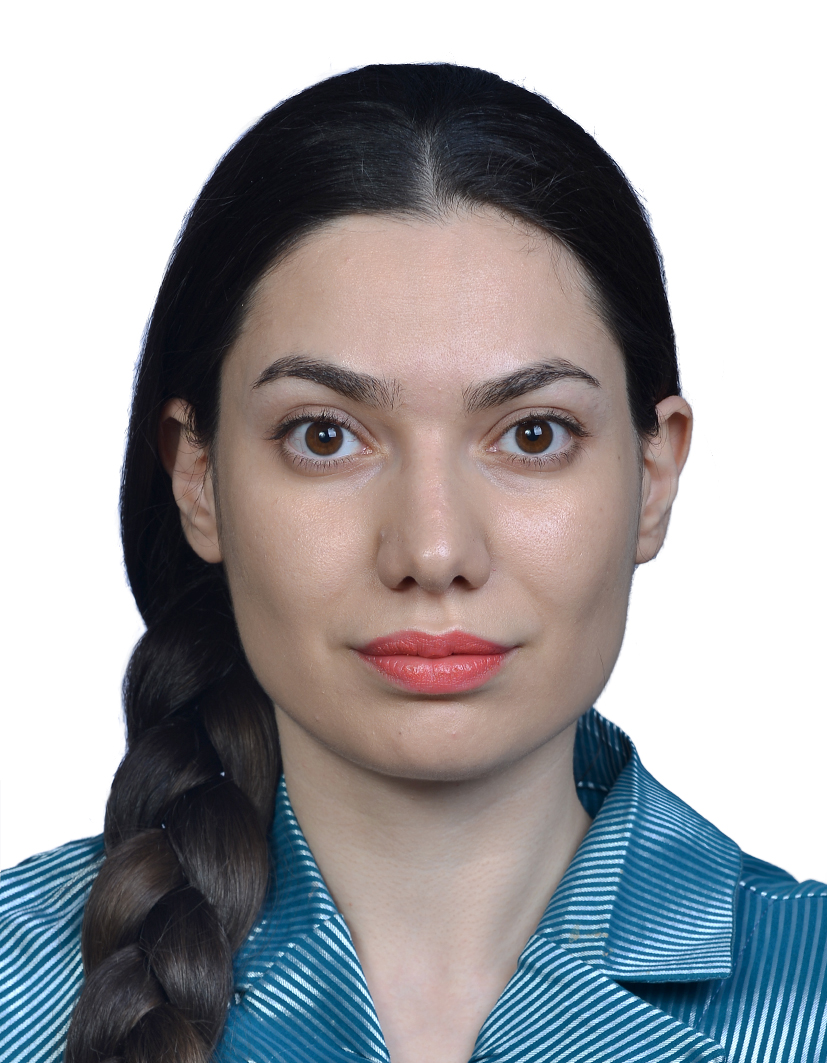}}]{Leila Sedghi} was born in Iran in 1990. She received B.S. and M.S. degrees in control engineering from Qazvin Islamic Azad University, Iran, in 2012 and 2016, respectively. She is currently pursuing a Ph.D. degree in computer science at University College Cork, Ireland, and her research is funded through the Confirm Centre for Smart Manufacturing.

She has been lecturing at the computer science department of Munster Technological University since 2019. Her research interests include control engineering, event-triggered control, machine learning, and deep learning. She has applied various control algorithms to different applications such as robotics, power systems, and microgrids.
\end{IEEEbiography}

\begin{IEEEbiography}[{\includegraphics[width=1in,height=1.25in,clip,keepaspectratio]{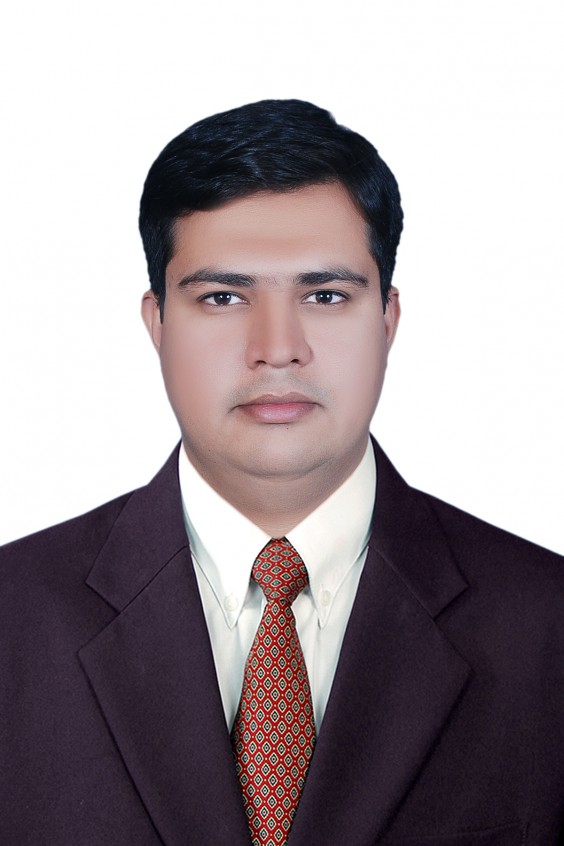}}]{Zohaib Ijaz} received the B.Sc. degree in Electronic Engineering from Islamia University of Bahawalpur, Pakistan in 2014 and Masters degree in Electrical Engineering with major in Control Systems from University of Engineering and Technology, Lahore, Pakistan in 2017. He started his Ph.D. in Computer Science at University College Cork, Ireland in 2019 funded through the Confirm Centre for Smart Manufacturing. His research interest is in Event based communication for Cyber-Physical Systems and consensus control of multi-agent systems.
\end{IEEEbiography}

\begin{IEEEbiography}[{\includegraphics[width=1in,height=1.25in,clip,keepaspectratio]{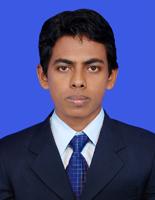}}]{Md. Noor-A-Rahim} received the Ph.D. degree from the Institute for Telecommunications Research, University of South Australia, Australia in 2015. He was a Postdoctoral Research Fellow with the Centre for Infocomm Technology (INFINITUS), Nanyang Technological University (NTU), Singapore. He is currently a Research Fellow with the School of Computer Science \& IT, University College Cork, Ireland. His research interests include Control over Wireless Networks, Intelligent Transportation Systems,  Machine Learning, Signal Processing, and DNA-based Data Storage. He was a recipient of the Michael Miller Medal from the Institute for Telecommunications Research (ITR), University of South Australia, for the most outstanding Ph.D. thesis in 2015.
\end{IEEEbiography}

\begin{IEEEbiography}[{\includegraphics[width=1in,height=1.25in,clip,keepaspectratio]{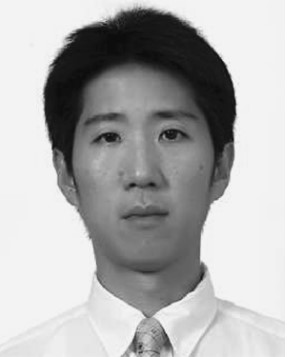}}]{Kritchai Witheephanich} received the M.Eng. degree in electrical engineering from King Mongkut's Institute of Technology Ladkrabang, Bangkok, Thailand, in 2001 and the Ph.D. degree in electronic and computer engineering from the University of Limerick, Limerick, Ireland, in 2012. Since 2001, he has been a Lecturer with the Department of Electrical Engineering, Srinakharinwirot University, Bangkok, where his research focuses on predictive control strategies to system science, control of communication networks, control of energy systems, and uncertain systems.
\end{IEEEbiography}

\begin{IEEEbiography}[{\includegraphics[width=1in,height=1.25in,clip,keepaspectratio]{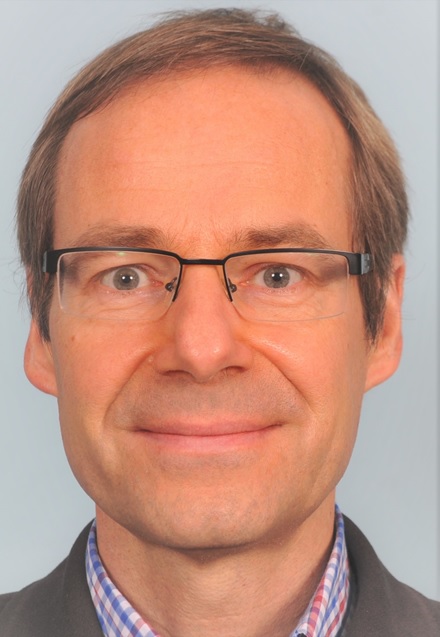}}]{Dirk Pesch} is a Professor in the School of Computer Science \& Information Technology at University College Cork and was previously Head of the Nimbus Research Centre at Cork Institute of Technology (now Munster Technological University). Dirk’s research interests focus on architecture, design, algorithms, and performance evaluation of low power, dense and vehicular wireless/mobile networks and services for Internet of Things and Cyber-Physical System’s applications in building management, smart connected communities, health and wellbeing, and smart manufacturing. He has over 25 years research and development experience in both industry and academia and has (co-)authored over 200 scientific articles and book chapters. He is a principal investigator of the national Science Foundation Ireland funded collaborative centres CONNECT (Future Networks) and CONFIRM (Smart Manufacturing), and director of the SFI Centre for Research Training in Advanced Networks for Sustainable Societies. He has also been involved in a number of EU funded research projects on smart and energy efficient buildings and urban neighbourhoods, including as coordinator. Dirk contributes to international conference organization and is a member of the editorial board of a number of journals. Dirk received a Dipl.Ing. degree from RWTH Aachen University, Germany, and a PhD from the University of Strathclyde, Glasgow, Scotland.
\end{IEEEbiography}

\EOD
\end{document}